\documentclass[journal,12pt]{IEEEtran}

\IEEEoverridecommandlockouts

\usepackage{cite}
\usepackage{algorithm}
\usepackage[noend]{algpseudocode}
\usepackage{amsmath,amssymb,amsfonts}
\usepackage[pdftex]{graphicx}
\usepackage{subcaption}
\usepackage{textcomp}
\usepackage{xcolor}
\usepackage{pifont}
\usepackage{multirow}
\usepackage{makecell}
\usepackage{array}
\usepackage{hyperref}

\begin{document}

\title{Deep Learning Enabled Semantic Communications with Speech Recognition and Synthesis}

\author{Zhenzi Weng, Zhijin Qin, Xiaoming Tao, Chengkang Pan, Guangyi Liu, \\and Geoffrey Ye Li
\thanks{Part of the work presented in~\cite{Weng2112:Semantic}, which has been published at IEEE GLOBECOM 2021.}
\thanks{Zhenzi Weng is with Queen Mary University of London, London E1 4NS, U.K. (e-mail: zhenzi.weng@qmul.ac.uk).}
\thanks{Zhijin Qin and Xiaoming Tao are with Tsinghua University, Beijing 100084, China (e-mail: qinzhijin@tsinghua.edu.cn; taoxm@tsinghua.edu.cn)(Corresponding author: Zhijin Qin).}
\thanks{Chengkang Pan and Guangyi Liu are with China Mobile Research Institute, Beijing, 100053, China (e-mail: panchengkang@chinamobile.com; liuguangyi@chinamobile.com).}
\thanks{Geoffrey Ye Li is with Imperial College London, London SW7 2AZ, U.K. (e-mail: geoffrey.li@imperial.ac.uk)}
\thanks{The work was supported in part by the National Natural Science Foundation of China (NSFC Nos 62293484 and 61925105) and in part by Tsinghua University-China Mobile Communications Group Company Ltd. Joint Institute.}
}

\maketitle
\begin{abstract}
In this paper, we develop a deep learning based semantic communication system for speech transmission, named DeepSC-ST. We take the speech recognition and speech synthesis as the transmission tasks of the communication system, respectively. First, the speech recognition-related semantic features are extracted for transmission by a joint semantic-channel encoder and the text is recovered at the receiver based on the received semantic features, which significantly reduces the required amount of data transmission without performance degradation. Then, we perform speech synthesis at the receiver, which dedicates to re-generate the speech signals by feeding the recognized text and the speaker information into a neural network module. To enable the DeepSC-ST adaptive to dynamic channel environments, we identify a robust model to cope with different channel conditions. According to the simulation results, the proposed DeepSC-ST significantly outperforms conventional communication systems and existing DL-enabled communication systems, especially in the low signal-to-noise ratio (SNR) regime. A software demonstration is further developed as a proof-of-concept of the DeepSC-ST.

\end{abstract}


\IEEEpeerreviewmaketitle

\section{Introduction}
With the booming of artificial intelligence (AI) in the recent years, the unprecedented demands on intelligent applications require extremely high transmission efficiency and impose enormous challenges on conventional communication systems. According to Shannon and Weaver~\cite{weaver1953recent}, the ultimate goal of communications is to exchange semantic information, named semantic communications. The transmission efficiency can be significantly improved with semantic communications.

Semantic information represents the meaning and veracity of source information~\cite{carnap1952outline}. However, it is challenging to quantify the semantic information due to the lack of a mathematical model. The breakthroughs of deep learning (DL) make it possible to tackle many challenges without requiring a mathematical model. Recently, DL-enabled semantic communications have shown great potentials to break the bottlenecks in conventional communication systems~\cite{qin2019deep} and facilitate semantic information exchange. Moreover, in some applications, users only request some critical semantic information from the source, which motivates us to transmit the application-related semantic information. Therefore, task-oriented semantic communications have been regarded as a promising solution for the six generation (6G) and beyond~\cite{qin2022semantic,9770094}, especially when communication resources are limited.

According to the state-of-the-art of DL-enabled semantic communications, the transmission goal could be categorized into two types: source data reconstruction and intelligent task execution. To achieve the data reconstruction, the global semantic information is extracted for transmission. But to serve the intelligent tasks, the extracted semantic information only consists of the task-related semantic features and the other irrelative features can be ignored to minimize the data to be transmitted. Besides, the semantic information can be significantly compressed by employing a lossless compression method~\cite{basu2014preserving}, which guarantees the feasibility to transmit the task-related semantic information instead of the global semantic information to achieve very high bandwidth efficiency.

Semantic communications have attracted intensive research for text~\cite{8461983,9398576,jiang2021deep,9700645,liang2022life}, speech/audio~~\cite{Weng2101:Semantic,tong2021federated,shi2021new}, and image/video transmission~\cite{9953076,9955991}. For semantic-aware speech/audio transmission, Weng~\emph{et al.}~\cite{Weng2101:Semantic} first proposed a DL-enabled semantic communication system, named DeepSC-S, to extract the global semantic information by leveraging an attention mechanism-powered module and reconstruct the speech signals at the receiver. Tong~\emph{et al.}~\cite{tong2021federated} developed a multi-user audio semantic communication system to collaboratively train the convolution neural network (CNN)-based autoencoder by implementing federated learning over multiple devices and a server. Moreover, Shi~\emph{et al.}~\cite{shi2021new} designed an understanding and transmission architecture for semantic communications and verified its effectiveness by deploying the architecture into the speech transmission system, which converts speech signals into semantic symbols to ensure high semantic fidelity and decodes the received semantic symbols into a speech waveform. Although semantic communications have been designed for intelligent speech transmission, investigation on semantic communications to execute speech-centric intelligent tasks at the receiver is still missing.

Due to the intensive deployment of intelligent devices in the post-Shannon communication era, a large amount of data transmission is inevitable to support massive connectivity amongst those devices. However, the available spectrum resources are scarce, which causes the bottleneck to the conventional communication system. Inspired by this, we propose a DL-enabled semantic communication system, named DeepSC-ST, for speech transmission and serving users with different requests. Particularly, DeepSC-ST first compresses the input speech sequence into the low-dimensional text-related semantic features that are transmitted over physical channels. At the receiver, the text sequence is estimated based on the received semantic features. By doing so, the characteristics of speech signals, e.g., the voice of speaker, speech delay, and background noise, etc., are omitted to be transmitted, which lowers the network traffic significantly and serves users requesting the text information only. Besides, to grant users the accessibility to speech signals at the receiver, the recognized text is passed through a speech synthesis module to restore the speech signals efficiently according to the user identity (ID). Note that the user ID is pre-registered and the corresponding speaker information is available at the receiver to reconstruct the speech sequence as close to the input speech sequence as possible. The main contributions of this paper are summarized as follows:
\begin{itemize}
\item A novel semantic communication system, named DeepSC-ST, is proposed for the communication scenarios with speech input, in which a joint semantic-channel coding scheme is developed.

\item The text-related semantic features are extracted from the input speech by leveraging CNN and recurrent neural network (RNN)-based semantic transmitter, which significantly reduces the transmission data and the required communication resources without performance degradation.

\item We develop speech recognition and speech synthesis tasks to achieve diverse system output. Particularly, the received text-related semantic features are converted to the text information by a feature decoder. Besides, the speech sequence is reconstructed according to the recovered text and the speaker information by leveraging a CNN and RNN-based neural network.

\item A demonstration of the DeepSC-ST with operable user interface is built to produce the recognized text and the synthesized speech based on the real human speech input.
\end{itemize}

The rest of this article is structured as follows. Section \uppercase\expandafter{\romannumeral2} presents the related work. The model of the semantic communication system for speech recognition and speech synthesis, as well as the performance metrics are introduced in Section \uppercase\expandafter{\romannumeral3}. In Section \uppercase\expandafter{\romannumeral4}, the proposed DeepSC-ST is detailed. Simulation results are discussed in Section \uppercase\expandafter{\romannumeral5} and Section \uppercase\expandafter{\romannumeral6} draws conclusions.

\emph{Notation}: The single boldface letters are used to represent vectors or matrices and single plain capital letters denote integers. $x_i$ indicates the $i$-th component of vector $\boldsymbol x$, $\left\|\boldsymbol x\right\|$ denotes the Euclidean norm of $\boldsymbol x$. $\boldsymbol Y\in\mathfrak R^{M\times N}$ indicates that $\boldsymbol Y$ is a $M\times N$ real matrix. Superscript swash letters refer the blocks in the system, e.g., $\mathcal T$ in $\boldsymbol\theta^{\mathcal T}$ represents the parameter at the transmitter. $\mathcal{CN}(\boldsymbol m,\;\boldsymbol V)$ denotes multivariate circular complex Gaussian distribution with mean vector $\boldsymbol m$ and co-variance matrix $\boldsymbol V$. Moreover, $\boldsymbol a\ast\boldsymbol b$ represents the convolution operation of vectors $\boldsymbol a$ and $\boldsymbol b$.

\section{Related Work}
In this section, we introduce the related work on DL-enabled semantic communication systems and present the state-of-the-art models for speech recognition and speech synthesis.

\subsection{Semantic Communication Systems}
Semantic communications have attracted extensive research interest very recently. Particularly, the transformer-powered system for text transmission, named DeepSC, has been proposed in~\cite{9398576} to measure the semantic error at the word level instead of the sentence level. Besides, a variant of DeepSC has been developed in~\cite{jiang2021deep} to further reduce the transmission error by leveraging hybrid automatic repeat request (HARQ) to improve the reliability of semantic transmission. Furthermore, the reasoning-based semantic communicator (R-SC) architecture in~\cite{liang2022life} automatically infers the hidden information by an inference function-based approach and introduces a life-long method to learn from previously received messages. In addition to perform text transmission, semantic communications for serving the text-based intelligent tasks have been investigated. Particularly, the transformer-enabled model, named DeepSC-MT in~\cite{9830752} can achieve the machine translation task by learning the word distribution of target language to map the meaning of source sentences to the target language. Besides, a visual question-answering (VQA) task has been investigated in~\cite{9830752} by designing a multi-modal system to compress the correlated text-image semantic features, and then to perform the information query at the receiver before fusing the text-image information to infer an accurate answer.

In semantic communications for image and video transmission, the DL-enabled semantic communication system for image transmission in~\cite{9953076} utilizes a generative adversarial network (GAN)-based semantic coding scheme to interpret the meaning of images and reconstruct images with high semantic fidelity. In~\cite{9955991}, a basal semantic video conference network has been established, which considers the impact of channel feedback and designs a semantic detector to detect semantic error by leveraging the ID classifier and fluency detection. Inspired by the booming of computer vision, semantic communications for serving intelligent vision tasks have shown great potentials to tackle many challenges beyond human limits. Particularly, the image classification task at the edge server has been investigated in~\cite{jankowski2020wireless}, which imposes IoT devices to compress images to the low computational complexity and to reduce the required transmission bandwidth. The semantic communication system for image classification in~\cite{9606667} adopts information bottleneck~\cite{tishby2000information} framework to identify the optimal tradeoff between compression rate and classification accuracy. In~\cite{9796572}, the deep reinforcement learning-enabled semantic communication system has been developed for joint image transmission and scene classification. Furthermore, a robust semantic communication system to combat semantic error has been proposed in~\cite{hu2022robust}, which incorporates the image with the generated semantic noise and utilizes the masked autoencoder to mitigate the effect of semantic noise with the aid of the discrete codebook shared by the transmitter and the receiver.

\subsection{Speech Recognition}
The exploration of speech recognition can be tracked back to several decades ago by exploiting the hidden Markov model~\cite{18626}. Afterwards, the neural network-powered speech recognition has experienced remarkable improvements owing to the thriving of natural language processing (NLP). About a decade ago, deep neural network (DNN) has been utilized for hybrid modeling of speech recognition system~\cite{5704567,5714717,6296526}, in which DNN replaces the traditional Gaussian mixture model that is utilized to estimate the distribution of hidden Markov model. However, such hybrid modeling architecture keeps all other models in the speech recognition system, i.e., acoustic model, lexicon model, and language model. Recently, a revolutionary transformation from hybrid modeling to end-to-end (E2E) modeling has been witnessed to directly recognize the token sequence from an input speech by leveraging a single integrated neural network, which simplifies the speech recognition pipeline and brings significant performance gains. Particularly, the speech recognition systems combining frequency-domain CNN with long short-term memory (LSTM) have been developed~\cite{6638947,pmlr-v48-amodei16}, which analyses the temporal dependencies of excessively long speech sequences and greatly increases the recognition accuracy compared to the hybrid modeling systems. The deep LSTM-based speech recognition systems have been proposed in~\cite{graves2014towards,sak14_interspeech,7472780} to perform character-level transcription and remove specific phonetic representation, which boosts the self-supervised learning technologies~\cite{baevski2020wav2vec,9054438,9414460}. Besides, RNN Transducer (RNN-T) has been utilized in E2E speech recognition systems due to its natural streaming capability and widely investigated in the academia and industry~\cite{8268937,8682336,9003906}. Furthermore, the attention-enabled speech recognition systems have been developed due to their capabilities on interaction amongst long sentences and superior training efficiency~\cite{9053896,gulati2020conformer}.

\subsection{Speech Synthesis}
The ultimate goal of speech synthesis is to generate intelligible and natural human speech corresponding to a text input, which implies the utterance of each word is correct, the intonation of synthesized speech must be similar to that of the native speaker, the speech quality should be good and free of any background noises or speech artifacts. The early speech synthesis systems involve a huge database of small sound units, which generates the speech waveform by concatenating many small sound units and arranges the order of these units by an appropriate algorithm\cite{moulines1990pitch,541110}. Inspired by the breakthroughs of DL, the speech synthesis systems employing different types of neural networks have been developed to generate very intelligible and clear speech\cite{6639215,7178816,7178814}. However, the speech produced by such neural network-based system still sounds mechanical and unnatural. The synthesized speech quality can be comparable to real human voice by generating waveform in the time domain from the input linguistic features since the advent of WaveNet~\cite{vandenoord16_ssw}. In~\cite{ping2017deep}, a convolutional attention-based speech synthesis system, Deep Voice 3~\cite{ping2017deep}, has been proposed, which is trained by the collected audio from over two thousand speakers. By taking text as inputs, Char2Wav~\cite{jose2017char2wav} generates speech waveform by utilizing the RNN-enabled reader and neural vocoder. The sequence-to-sequence architecture, named Tacotron~\cite{Wang2017TacotronTE}, maps the text into magnitude spectrogram instead of linguistic and acoustic features, and simplifies the conventional synthesis procedure by a single neural network trained separately. The improved system, Tacotron 2~\cite{8461368}, produces the speech waveform from normalized character sequences, which synthesizes the realistic human voice. Although tremendous success of the above autoregressive speech synthesis systems, the time to generate speech with long sentences could be several seconds. More recently, some non-autoregressive speech synthesis pipelines have been developed~\cite{9054484,donahue2020end,ren2020fastspeech} to eliminate the time dependency of the produced waveform and reduce the latency of the synthesis process.
\section{System Model}
The semantic communication system for speech recognition and speech synthesis first extracts and transmits the low-dimensional text-related semantic features from the input speech, then recognizes the text sequence based on the received semantic features. Finally, the speech waveform is reconstructed at the receiver according to the recognized text and the user ID. In this section, we introduce the considered system model and performance metrics for speech recognition and speech synthesis tasks.

\subsection{Input Spectrum and Text Information}
The input speech sample sequence is converted into the spectrum before feeding into the transmitter. First, the input speech sample sequence, $\boldsymbol m=\left[m_1,\boldsymbol\;m_2,\;\dots,\;m_Q\right]$, is divided into $N$ frames, then these frames are converted into the spectrum through the Hamming window, fast Fourier transform (FFT), logarithm operation, and normalization. By doing so, the spectrum, $\boldsymbol s=\left[{\boldsymbol s}_{\mathbf1},\boldsymbol\;{\boldsymbol s}_{\mathbf2},\;\dots,\;{\boldsymbol s}_{\mathbf N}\right]$, contains the characteristics of the sample sequence, $\boldsymbol m$.

Moreover, denote $\boldsymbol t$ as the corresponding text of the single speech sample sequence, $\boldsymbol m$. The ultimate goal of the speech recognition task is to recover the final text transcription, $\widehat{\boldsymbol t}$, as close to $\boldsymbol t$ as possible. Denote $\boldsymbol t=\left[t_1,\boldsymbol\;t_2,\;\dots,\;t_K\right]$, where $t_k$ is a token from the token set, $\boldsymbol t$, that could be a character in the alphabet or a word boundary. For English, there are 26 characters in the alphabet. Then there are 29 tokens if including $apostrophe$, $space$, and $blank$ as word boundaries, that is, $\overline{\boldsymbol t}=\left[\mathrm a,\;\mathrm b,\;\mathrm c,\;\dots,\;\mathrm z,\;apostrophe,\;space,\;blank\right]$.

\subsection{Transmitter}
\begin{figure*}[tbp]
\centering 
\includegraphics[width=0.9\textwidth]{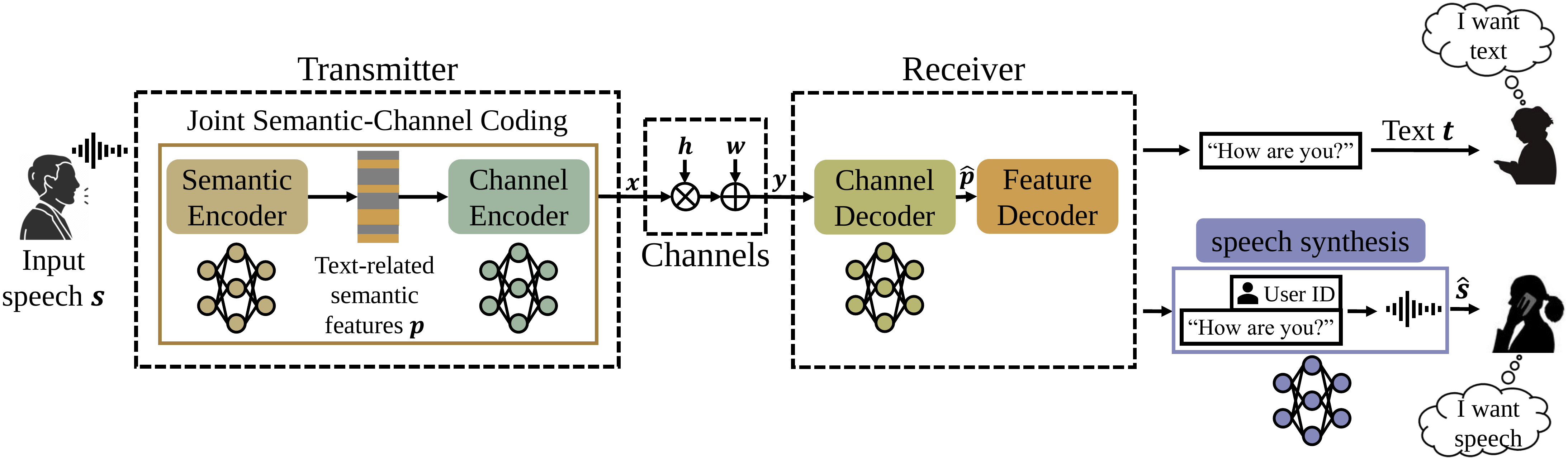} 
\caption{The proposed DL-enabled semantic communication system for speech recognition and speech synthesis (DeepSC-ST).}
\label{sys model}
\end{figure*}

Based on the input spectrum and text sequence of the speech sample sequence, the proposed system model is shown in Fig.~\ref{sys model}. In the figure, the transmitter consists of the~\emph{semantic encoder} and the~\emph{channel encoder}, implemented by two neural networks. At the transmitter, the input spectrum, $\boldsymbol s$, is converted into the text-related semantic features, $\boldsymbol p$, by the~\emph{semantic encoder}, and these features are mapped into symbols, $\boldsymbol x$, by the~\emph{channel encoder} to be transmitted over physical channels. Denote the neural network parameters of the~\emph{semantic encoder} and the~\emph{channel encoder} as $\boldsymbol\alpha$ and $\boldsymbol\beta$, respectively, then the neural network parameters at the transmitter can be expressed as $\boldsymbol\theta^{\mathcal T}=(\boldsymbol\alpha,\boldsymbol\;\boldsymbol\beta)$. Hence, the encoded symbols, $\boldsymbol x$, can be expressed as
\begin{equation}
\boldsymbol x=\mathbf T_{\boldsymbol\beta}^{\mathcal C}(\mathbf T_{\boldsymbol\alpha}^{\mathcal S}(\boldsymbol s)),
\label{auto-encoder}
\end{equation}
where $\mathbf T_{\boldsymbol\alpha}^{\mathcal S}(\cdot)$ and $\mathbf T_{\boldsymbol\beta}^{\mathcal C}(\cdot)$ indicate the~\emph{semantic encoder} and the~\emph{channel encoder} with respect to (w.r.t.) parameters $\boldsymbol\alpha$ and $\boldsymbol\beta$, respectively.

The encoded symbols, $\boldsymbol x$, are transmitted over a physical channel. $\boldsymbol x$ is assumed to be normalized, i.e., $\mathbb{E}\left\|\boldsymbol x\right\|^2=1$. 

In Fig.~\ref{sys model}, the wireless channel, represented by $p_h\left(\left.\boldsymbol y\right|\boldsymbol x\right)$, takes $\boldsymbol x$ as the input and produces the output, $\boldsymbol y$, as the received symbols. Denote the coefficients of a linear channel as $\boldsymbol h$, then the transmission process from the transmitter to the receiver can be modeled as
\begin{equation}
\boldsymbol y=\boldsymbol h\ast\boldsymbol x+\boldsymbol w,
\label{channel}
\end{equation}
where $\boldsymbol w\sim\mathcal{CN}(0,\;\sigma^2\mathbf I)$ denotes independent and identically distributed (i.i.d.) Gaussian noise vector with variance $\sigma^2$ for each channel.

\subsection{Receiver}
As in Fig.~\ref{sys model}, the receiver includes the~\emph{channel decoder} and the~\emph{feature decoder} to recover the text-related semantic features and recognize the final text transcription as close to the raw text sequence as possible. First, the received symbols, $\boldsymbol y$, is mapped into the text-related semantic features, $\widehat{\boldsymbol p}$, by the~\emph{channel decoder}, where $\widehat{\boldsymbol p}=\left[{\widehat{\boldsymbol p}}_1,\;{\widehat{\boldsymbol p}}_2,\;\dots,\;{\widehat{\boldsymbol p}}_L\right]$ denotes a probability matrix and probability vector ${\widehat{\boldsymbol p}}_l=\left[\widehat p_l^1,\;\widehat p_l^2,\;\dots,\;\widehat p_l^{29}\right]$ comprises 29 probabilities corresponding to 29 tokens in $\overline{\boldsymbol t}$. Denote the neural network parameters of the~\emph{channel decoder} as $\boldsymbol\theta^{\mathcal R}$, then the recovered features, $\widehat{\boldsymbol p}$, can be obtained from the received symbols, $\boldsymbol y$, by
\begin{equation}
\widehat{\boldsymbol p}=\mathbf R_{\boldsymbol\theta^{\mathcal R}}^{\mathcal S}(\boldsymbol y),
\label{auto-decoder}
\end{equation}
where $\mathbf R_{\boldsymbol\theta^{\mathcal R}}^{\mathcal S}(\cdot)$ indicates the~\emph{channel decoder} w.r.t. parameters $\boldsymbol\theta^{\mathcal R}$.

Then, the text-related semantic features, $\widehat{\boldsymbol p}$, are decoded into the text transcription, $\widehat{\boldsymbol t}$, by the~\emph{feature decoder}, denoted as
\begin{equation}
\widehat{\boldsymbol t}=\mathbf R^{\mathcal F}(\widehat{\boldsymbol p}),
\label{feature decoder}
\end{equation}
where $\mathbf R^{\mathcal F}(\cdot)$ represents the~\emph{feature decoder}.

The objective of the semantic communication system for speech recognition task is to recover the text information of the input speech signals, which is equivalent to maximizing the posterior probability $p\left(\left.\boldsymbol t\right|\boldsymbol s\right)$. By introducing connectionist temporal classification (CTC)~\cite{graves2006connectionist}, the posterior probability $p\left(\left.\boldsymbol t\right|\boldsymbol s\right)$ can be expressed as
\begin{equation}
p\left(\left.\boldsymbol t\right|\boldsymbol s\right)=\sum_{A\in\mathfrak A(\boldsymbol s,\boldsymbol\;\boldsymbol t)}\left(\prod_{l=1}^L{\widehat p}_l\left(\left.a_l\right|\boldsymbol s\right)\right),
\label{posterior probability CTC}
\end{equation}
where $\mathfrak A(\boldsymbol s,\boldsymbol\;\boldsymbol t)$ represents the set of all possible valid alignments of text sequence $\boldsymbol t$ to spectrum $\boldsymbol s$, and $a_l$ is the token under the valid alignments. For example, if text sequence $\boldsymbol t=\left[\mathrm t,\;\mathrm a,\;\mathrm s,\;\mathrm t,\;\mathrm e\right]$, the valid alignments could be $\left[blank,\;\mathrm t,\;blank,\;\mathrm a,\;\mathrm s,\;blank,\;\mathrm t,\;\mathrm e\right]$, or $\left[\mathrm t,\;blank,\;\mathrm a,\;\mathrm s,\;blank,\;blank,\;\mathrm t,\;\mathrm e\right]$, etc., because the $blank$ token is removed when obtaining the final text transcription, $\widehat{\boldsymbol t}$. Note that the number of tokens in every valid alignment is $L$. If the valid alignment is $\left[blank,\;\mathrm t,\;blank,\;\mathrm a,\;\mathrm s,\;blank,\;\mathrm t,\;\mathrm e\right]$, the first token is $blank$, i.e., $a_1=blank$, then we have
\begin{equation}
{\widehat p}_l\left(\left.a_l\right|\boldsymbol s\right)={\widehat p}_l\left(\left.blank\right|\boldsymbol s\right)=\widehat p_l^{29},\;\;\;\;\;l=1,
\label{probability pl}
\end{equation}
where $\widehat p_l^{29}$ is one of the probabilities in probability vector ${\widehat{\boldsymbol p}}_l$ and number 29 represents the $blank$ token is the 29th token in $\overline{\boldsymbol t}$.

To maximize the posterior probability $p\left(\left.\boldsymbol t\right|\boldsymbol s\right)$, the CTC loss is adopted as the loss function for speech recognition task in our system, denoted as
\begin{equation}
{\mathcal L}_{CTC}(\boldsymbol\theta)=-\ln\left(\sum_{A\in\mathfrak A(\boldsymbol s,\boldsymbol\;\boldsymbol t)}\left(\prod_{l=1}^L{\widehat p}_l\left(\left.a_l\right|\boldsymbol s,\boldsymbol\theta\right)\right)\right),
\label{CTC loss}
\end{equation}
where $\boldsymbol\theta$ denotes the neural network parameters of the transmitter and the receiver, $\boldsymbol\theta=(\boldsymbol\theta^{\mathcal T},\;\boldsymbol\theta^{\mathcal R})$. 

Moreover, for given prior channel state information (CSI), the neural network parameters, $\boldsymbol\theta$, can be updated by the stochastic gradient descent (SGD) algorithm as follows,
\begin{equation}
\boldsymbol\theta^{(i+1)}\leftarrow\boldsymbol\theta^{(i)}-\eta\nabla_{\boldsymbol\theta^{(i)}}{\mathcal L}_{CTC}(\boldsymbol\theta),
\label{SGD}
\end{equation}
where $\eta>0$ is a learning rate and $\nabla$ indicates the nabla operator.

\subsection{Speech Synthesis Module}
By introducing a flexible task mechanism, the produced information at the receiver is not only limited to the text transcription, but also the speech information, which grants users the privilege to check the speech characteristics and expands the system diversity. As shown in Fig.~\ref{sys model}, the input to the~\emph{speech synthesis module} refers to the output of the~\emph{feature decoder}, $\widehat{\boldsymbol t}$, and the~\emph{speech synthesis module}, represented by a neural network, processes and converts $\widehat{\boldsymbol t}$ into the speech sample sequence, $\widehat{\boldsymbol m}$, which dedicates to reconstruct the speech waveform as close to the original sample sequence, $\boldsymbol m$, as possible. Denote the neural network parameters of the~\emph{speech synthesis module} as ${\boldsymbol\chi}$, then the reconstructed speech sample sequence, $\widehat{\boldsymbol m}$, can be denoted as
\begin{equation}
\widehat{\boldsymbol m}=\mathbf R_{\boldsymbol\chi}^{\mathcal{SS}}\left(\widehat{\boldsymbol t}\right),
\label{speech synthesis module}
\end{equation}
where $\mathbf R_{\boldsymbol\chi}^{\mathcal{SS}}\left(\cdot\right)$ indicates the~\emph{speech synthesis module} w.r.t. parameters $\boldsymbol\chi$.

\subsection{Performance Metrics}
\subsubsection{Speech Recognition Task}
In this task, the semantic similarity between the recovered text and the original text is equivalent to measuring whether the recovered text is as readable and understandable as the original text. It is intuitive that the text sequence is easier to read and understand if it contains fewer incorrect characters/words, which inspires the metrics by calculating the incorrect characters/words in the recovered text sequence. Besides, thanks to the advanced developments in speech recognition field, character error-rate (CER) and word error-rate (WER) are effective metrics to indicate the accuracy of the recognized text transcription. Therefore, we adopt CER and WER as two performance metrics for the speech recognition task. According to the text transcription, $\widehat{\boldsymbol t}$, the substitution, deletion, and insertion operations in character are utilized to restore the raw text sequence, $\boldsymbol t$. The calculation of CER can be denoted as
\begin{equation}
\mathrm{CER}=\frac{S_C+D_C+I_C}{N_C},
\label{CER}
\end{equation}
where $S_C$, $D_C$, and $I_C$ represent the numbers of character substations, deletions, and insertions, respectively, and $N_C$ is the number of characters in $\boldsymbol t$.

Similarly, the substitution, deletion, and insertion operations in word are employed to calculate WER, which can be expressed as
\begin{equation}
\mathrm{WER}=\frac{S_W+D_W+I_W}{N_W},
\label{WER}
\end{equation}
where $S_W$, $D_W$, and $I_W$ denote the numbers of word substations, deletions, and insertions, respectively, and $N_W$ is the number of words in $\boldsymbol t$.

Note that CER and WER may exceed one owing to a large number of deletions. Moreover, for the same sentence, CER is typically lower than WER. In reality, the recognized sentence is usually readable when CER is lower than around 0.15.

\subsubsection{Speech Synthesis Task}
The ultimate goal of the speech synthesis task is to reconstruct the clear speech. Due to the difficulty to attach audio to the paper for listening, the reasonable approach for assessing the quality of the synthesized speech is to compare it with the real speech. In this task, we adopt unconditional Fréchet deep speech distance (FDSD) and unconditional kernel deep speech distance (KDSD)~\cite{binkowski2019high} as two quantitative metrics to evaluate the distribution similarity between the synthesized speech and the real speech by measuring the Fréchet distance and the maximum mean discrepancy (MMD) between them, respectively. The lower the FDSD or KDSD values, the higher similarity between the synthesized speech and the real speech, i.e., the higher quality of the synthesized speech.

Given the real speech sample sequence, $\boldsymbol m$, the synthesized speech sample sequence, $\widehat{\boldsymbol m}$, and a publicly available deep speech recognition model, the features of real and synthesized speech sample sequences, denoted as $\boldsymbol D\in\mathfrak R^{U\times V}$ and $\widehat{\boldsymbol D}\in\mathfrak R^{\widehat U\times V}$ respectively, can be extracted by passing $\boldsymbol m$ and $\widehat{\boldsymbol m}$ through the speech recognition model, respectively. Therefore, their FDSD can be calculated as
\begin{equation}
\begin{split}
    \mathrm{FDSD}^2&=\left\|{\boldsymbol\mu}_{\boldsymbol D}-{\boldsymbol\mu}_{\widehat{\boldsymbol D}}\right\|^2 \\
    &+\mathrm{Tr}\left[{\boldsymbol\Sigma}_{\boldsymbol D}+{\boldsymbol\Sigma}_{\widehat{\boldsymbol D}}-\left({\boldsymbol\Sigma}_{\boldsymbol D}{\boldsymbol\Sigma}_{\widehat{\boldsymbol D}}\right)^\frac12\right],
\end{split}
\label{cFDSD}
\end{equation}
where ${\boldsymbol\mu}_{\boldsymbol D}$ and ${\boldsymbol\mu}_{\widehat{\boldsymbol D}}$ represent the means of $\boldsymbol D$ and $\widehat{\boldsymbol D}$, respectively, ${\boldsymbol\Sigma}_{\boldsymbol D}$ and ${\boldsymbol\Sigma}_{\widehat{\boldsymbol D}}$ denote their covariance matrices, and $\mathrm{Tr}(\cdot)$ indicates the trace of a square matrix.

On the other hand, KDSD can be obtained by~(\ref{cKDSD}),
\begin{equation}
\begin{split}
    \mathrm{KDSD}^2&=\frac1{U\left(U-1\right)}\sum_{\underset{i\neq j}{1\leq i,j\leq U}}kf\left({\boldsymbol D}_{i,}{\widehat{\boldsymbol D}}_j\right) \\
    &+\frac1{\widehat U\left(\widehat U-1\right)}\sum_{\underset{i\neq j}{1\leq i,j\leq\widehat U}}kf\left({\boldsymbol D}_{i,}{\widehat{\boldsymbol D}}_j\right) \\
    &+\sum_{i=1}^U\sum_{j=1}^{\widehat U}kf\left({\boldsymbol D}_{i,}{\widehat{\boldsymbol D}}_j\right),
\end{split}
\label{cKDSD}
\end{equation}
where $kf(\cdot)$ is a kernel function, defined as
\begin{equation}
kf\left({\boldsymbol D}_{i,}{\widehat{\boldsymbol D}}_j\right)=\left(\frac1V{\boldsymbol D}_i{\widehat{\boldsymbol D}_j}^T+1\right)^3.
\label{kernel function}
\end{equation}

\section{Semantic Communications for Speech Recognition and Synthesis}
In this section, we present the details of the DeepSC-ST. Specifically, in the speech recognition task, CNN and RNN are adopted for the semantic encoding, dense layers are employed for the channel encoding and decoding. In the speech synthesis task, the Tacotron 2~\cite{8461368} is utilized to reconstruct the speech waveform from the recognized text according to the user ID.

\subsection{Model Description}
\begin{figure*}[tbp]
\includegraphics[width=1.0\textwidth]{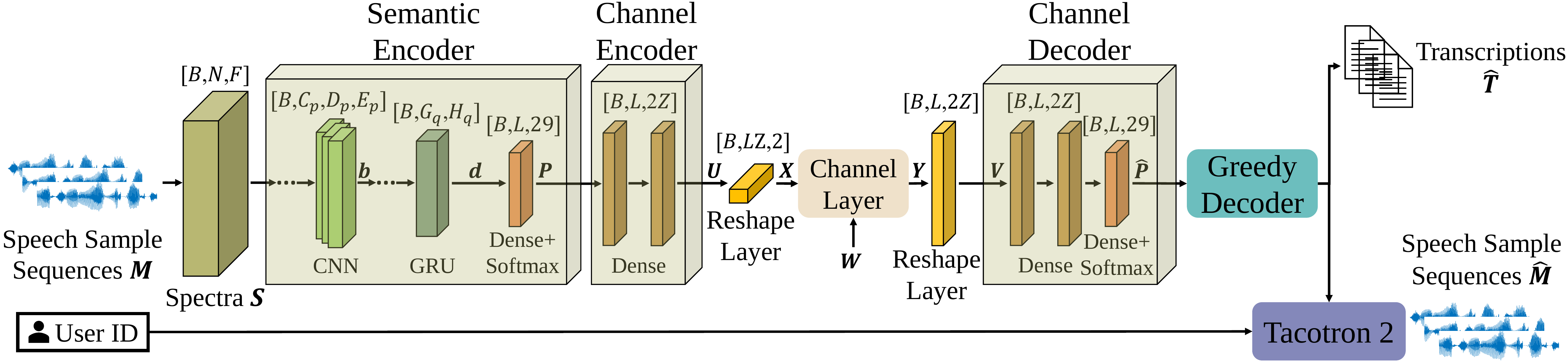} 
\centering 
\caption{The proposed system architecture for semantic communication system for speech recognition and speech synthesis.}  
\label{proposed sys}
\end{figure*}

The proposed DeepSC-ST is shown in Fig.~\ref{proposed sys}. In the figure, $\boldsymbol M$ is the set of speech sample sequences drawn from the speech dataset and is converted into a set of spectra, $\boldsymbol S=\left[{\boldsymbol S}_1,\;{\boldsymbol S}_2,\;...,\;{\boldsymbol S}_B\right]$, where $B$ is the batch size. In addition, $\boldsymbol T$ is the set of correct text sequences, $\boldsymbol t$, corresponding to $\boldsymbol M$. The spectra, $\boldsymbol S$, are fed into the~\emph{semantic encoder} to learn and extract the text-related semantic features and to output the features, $\boldsymbol P$. The details of the~\emph{semantic encoder} are presented in part B of this section. Afterwards, the~\emph{channel encoder}, implemented by two dense layers, converts $\boldsymbol P$ into $\boldsymbol U$. To transmit $\boldsymbol U$ into a physical channel, it is reshaped into symbols, $\boldsymbol X$, via a reshape layer.

The received symbols, $\boldsymbol Y$, are reshaped into $\boldsymbol V$ before feeding into the~\emph{channel decoder}, represented by three dense layers. The output of the~\emph{channel decoder} is the recovered text-related semantic features, $\widehat{\boldsymbol P}$. Next, the~\emph{greedy decoder}, i.e., the~\emph{feature decoder}, decodes $\widehat{\boldsymbol P}$ into the text transcriptions, $\widehat{\boldsymbol T}$. The details of the~\emph{greedy decoder} are presented in part C of this section.

Furthermore, from the Fig.~\ref{proposed sys}, the speech sample sequences, $\widehat{\boldsymbol M}$, can be reconstructed by the~\emph{speech synthesis module}, i.e., Tacotron 2, by combing the corresponding user ID. It is worth mentioning that Tacotron 2 is trained separately from the speech recognition task and can be omitted in the communication scenarios where users only request the text information. Some details of Tacotron 2 are introduced in part D of this section.

\subsection{Semantic Encoder}
The~\emph{semantic encoder} is constructed by the CNN and the gated recurrent unit (GRU)-based bidirectional RNN (BRNN)~\cite{650093} modules. As shown in Fig.~\ref{proposed sys}, the input spectra, $\boldsymbol S$, are first converted into the intermediate features via several CNN modules. Particularly, the number of~\emph{filters} in each CNN module is $E_p$, $p\in\left[1,\;2,\;\dots,\;P\right]$, and the output of the last CNN module is $\boldsymbol b\in\mathfrak R^{B\times C_P\times D_P\times E_P}$. Then, $\boldsymbol b$ is fed into $Q$ BRNN modules, successively, and produces $\boldsymbol d\in\mathfrak R^{B\times G_Q\times H_Q}$, where the number of GRU~\emph{units} in each BRNN module, $H_q$, $q\in\left[1,\;2,\;\dots,\;Q\right]$, is consistent. Finally, the text-related semantic features, $\boldsymbol P$, are obtained from $\boldsymbol d$ by passing through multiple cascaded dense layers and a softmax layer.

\subsection{Greedy Decoder}
\begin{figure}[tbp]
\includegraphics[width=0.45\textwidth]{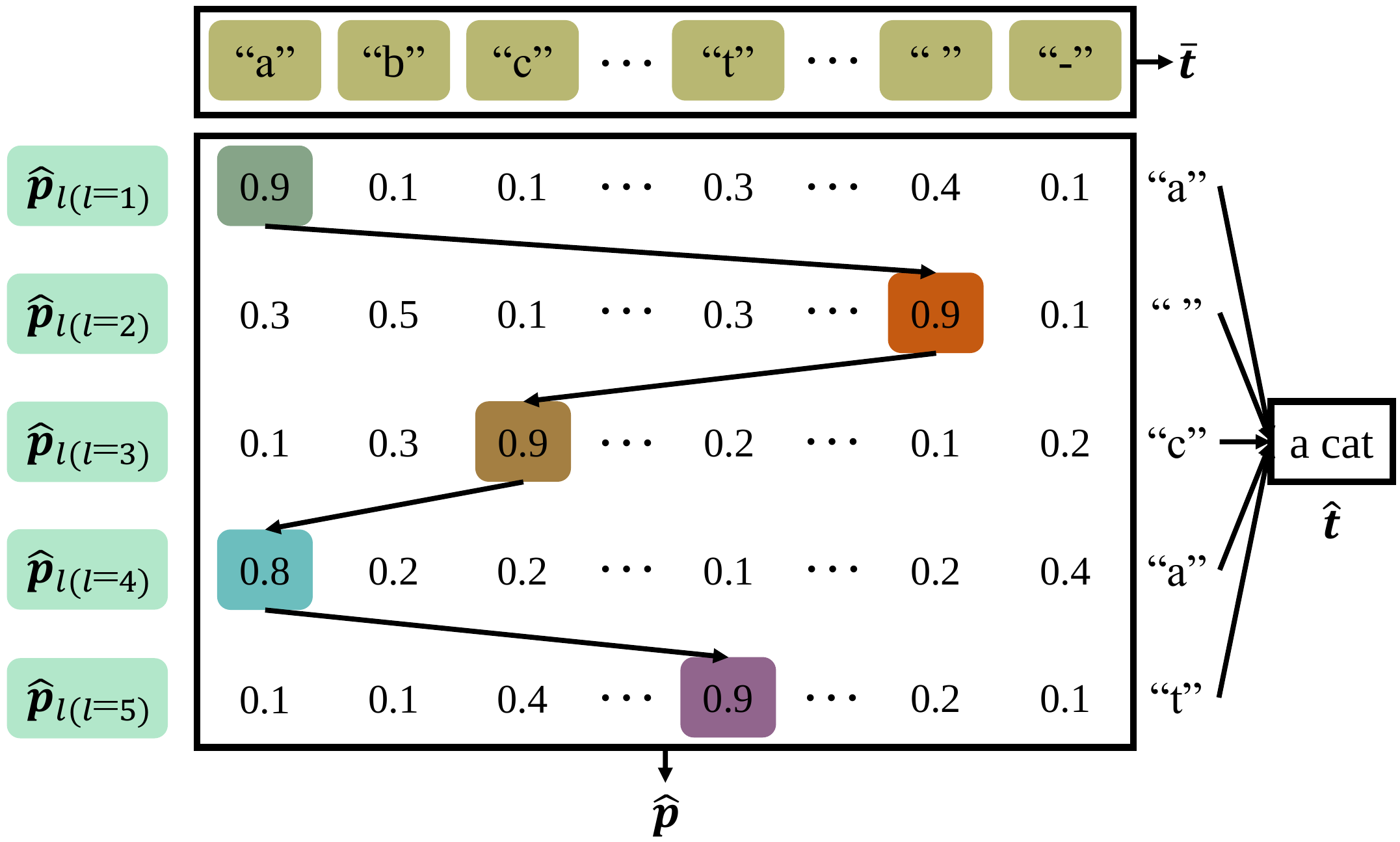} 
\centering 
\caption{An example of the~\emph{greedy decoder}.}  
\label{greedy decoder}
\end{figure}

As aforementioned that the recovered features, $\widehat{\boldsymbol P}$, are decoded into the text transcriptions, $\widehat{\boldsymbol T}$, via the~\emph{greedy decoder}. An example to obtain the text transcription by the~\emph{greedy decoder} is shown in Fig.~\ref{greedy decoder}. During the decoding process, in each step $l$, the maximum probability in the vector, ${\widehat{\boldsymbol p}}_l$, is selected and mapped to the corresponding token in $\overline{\boldsymbol t}$. The final text transcription, $\widehat{\boldsymbol t}$ is composed by concatenating all mapped tokens.

It is worth mentioning that the processes of selecting the maximum probability in ${\widehat{\boldsymbol p}}_l$ and mapping the maximum probability to the corresponding token in $\overline{\boldsymbol t}$ are non-differentiable, which runs counter to the prerequisite of a differentiable loss function to design a neural network. Therefore, the~\emph{greedy decoder} is unable to be implemented by the neural network.

\subsection{Training and Testing for Speech Recognition Task}
According to the prior knowledge of CSI, the training and testing algorithms for speech recognition task are described in Algorithm~\ref{training algorithm} and Algorithm~\ref{testing algorithm}, respectively. During the training stage, the~\emph{greedy decoder} and the~\emph{speech synthesis module} are omitted.
\begin{algorithm}[tbp]
\caption{Training algorithm for speech recognition task.}
\label{training algorithm}
\textbf{Initialization:} initialize parameters $\boldsymbol\theta^{(0)}$, $i=0$.

\begin{algorithmic}[1]
    \State \textbf{Input:} Speech sample sequences $\boldsymbol M$ and transcriptions $\boldsymbol T$ from trainset $\mathfrak S$, fading channel $\boldsymbol H$, noise $\boldsymbol W$.
    \State Generate spectra $\boldsymbol S$ from sample sequences $\boldsymbol M$.
        \While{CTC loss is not converged}
            \State $\mathbf T_{\boldsymbol\beta}^{\mathcal C}(\mathbf T_{\boldsymbol\alpha}^{\mathcal S}(\boldsymbol S))\rightarrow\boldsymbol X$.
            \State Transmit $\boldsymbol X$ and receive $\boldsymbol Y$ via (\ref{channel}).
            \State $\mathbf R_{\boldsymbol\delta}^{\mathcal S}(\boldsymbol Y)\rightarrow\widehat{\boldsymbol P}$.
            \State Compute loss ${\mathcal L}_{CTC}(\boldsymbol\theta)$ via (\ref{CTC loss}).
            \State Update parameters $\boldsymbol\theta$ via SGD according to (\ref{SGD}).
            \State $i\leftarrow i+1$.
        \EndWhile
    \State \textbf{end while}
    \State \textbf{Output:} Trained networks $\mathbf T_{\boldsymbol\alpha}^{\mathcal S}(\cdot)$, $\mathbf T_{\boldsymbol\beta}^{\mathcal C}(\cdot)$, and $\mathbf R_{\boldsymbol\delta}^{\mathcal C}(\cdot)$.
\end{algorithmic}

\end{algorithm}
\begin{algorithm}[tbp]
\caption{Testing algorithm for speech recognition task.}
\label{testing algorithm}

\begin{algorithmic}[1]   
    \State \textbf{Input:} Speech sample sequences $\boldsymbol M$ from testset, trained networks $\mathbf T_{\boldsymbol\alpha}^{\mathcal S}(\cdot)$, $\mathbf T_{\boldsymbol\beta}^{\mathcal C}(\cdot)$, and $\mathbf R_{\boldsymbol\delta}^{\mathcal C}(\cdot)$, testing channel set $\mathcal H$, a wide range of SNR regime.
    \State Generate spectra $\boldsymbol S$ from sample sequences $\boldsymbol M$.
    	\For{channel condition $\boldsymbol H$ drawn from $\mathcal H$}
    	    \For{each SNR value}
    	        \State Generate Gaussian noise $\boldsymbol W$ under the SNR value.
    	        \State $\mathbf T_{\boldsymbol\beta}^{\mathcal C}(\mathbf T_{\boldsymbol\alpha}^{\mathcal S}(\boldsymbol S))\rightarrow\boldsymbol X$.
                \State Transmit $\boldsymbol X$ and receive $\boldsymbol Y$ via (\ref{channel}).
                \State $\mathbf R_{\boldsymbol\delta}^{\mathcal S}(\boldsymbol Y)\rightarrow\widehat{\boldsymbol P}$.
                \State Decoding $\widehat{\boldsymbol P}$ into $\widehat{\boldsymbol T}$ via (\ref{feature decoder}).
                \EndFor
            \State \textbf{end for}
        \EndFor
    \State \textbf{end for}
	\State \textbf{Output:} Recovered text transcriptions, $\widehat{\boldsymbol S}$.
\end{algorithmic}

\end{algorithm}

\subsection{Speech Synthesis Module}\label{speech synthesis details}
As shown in Fig.~\ref{proposed sys}, Tacotron 2 is adopted to reconstruct the speech sample sequences, $\widehat{\boldsymbol M}$, as close to the input sample sequences, $\boldsymbol M$, as possible. Tacotron 2 is composed of the spectrogram prediction network and a variant of WaveNet vocoder~\cite{tamamori2017speaker}, which converts the input token sequence into the mel-frequency spectrogram and produces time-domain waveform according to the predicted spectrogram, respectively. The mel-frequency spectrogram is obtained by implementing a nonlinear process to the frequency of the short-time Fourier transform (STFT) magnitude, which is straightforward for the WaveNet model to generate audio owing to the simple and low-level acoustic characteristic. Particularly, an encoder maps the input token sequence into the internal feature representation after a 512-dimensional token embedding, which is achieved by a stack of three convolutional layers with 512~\emph{filters} followed by batch normalization and ReLU activation, as well as a bidirectional long short-term memory (LSTM) layer containing 512~\emph{units}.

As a consequence, the encoded features are processed by a decoder enabled by a neural network to predict the mel-frequency spectrogram frame by frame. In details, the previous synthesized frame is fed into a~\emph{pre-net} with two connected layers of 256 ReLU~\emph{units}, then the output is combined with the encoded features and the user ID to predict the current frame after passing through a stack of two LSTM layers containing 1024~\emph{units} and a~\emph{post-net} with a stack of five convolutional layers with 512~\emph{filters} followed by batch normalization and ReLU activation. The spectrogram prediction network is trained independently by minimizing the sum of the mean-squared error (MSE) before and after the~\emph{post-net}.

Furthermore, the mel-frequency spectrogram frames are fed into the WaveNet vocoder to predict the parameters, e.g., mean and log scale, of the synthesized waveform after a ReLU activation followed by a linear projection, which encourages to train the WaveNet vocoder by maximizing the log-likelihood of the speech waveform w.r.t. the trainable parameters due to the tractability of log-likelihoods.

\section{Numerical Results}
In this section, we compare the proposed DeepSC-ST with the conventional communication systems and the existing semantic communication systems under different channels, where the accurate CSI is assumed at the receiver. The experiment is conducted on the LJSpeech dataset~\cite{ljspeech17} for both speech recognition and speech synthesis tasks, which is a corpus of English speech with the sampling rate of 22,050 Hz. The speech is down-sampled into 16,000 Hz. The adopted simulation environment is Tensorflow 2.4.

\subsection{Simulation Setting and Benchmarks}
In the proposed DeepSC-ST, the numbers of CNN modules and BRNN modules in the~\emph{semantic encoder} are two and six, respectively. The number of~\emph{filters} for each CNN module is 32 and the number of GRU~\emph{units} for each BRNN module is 800. Moreover, two dense layers are utilized in the~\emph{channel encoder} with 40~\emph{units} and three dense layers are utilized in the~\emph{channel decoder} with 40, 40, and 29~\emph{units}, respectively. The batch size is $B=24$ and the learning rate is $\eta=0.0001$. The parameter settings of the proposed DeepSC-ST for speech recognition task are summarized in Table~\ref{DeepSC-ST NN parameters}. For performance comparison, we provide the following four benchmarks:
\subsubsection{\textbf{Benchmark 1}}
The first benchmark is a conventional system that transmits speech signals, named speech transceiver. The input of the system is the speech signals, which is restored at the receiver. Moreover, the text transcription is obtained from the recovered speech signals after passing through a speech recognition model. The adaptive multi-rate wideband (AMR-WB) system~\cite{1175533} is used for speech source coding and 64-QAM is utilized for modulation. Polar code with successive cancellation list (SCL) decoding algorithm~\cite{7114328} is employed for channel coding with the block length of 512 bits and the list size of four. Moreover, the speech recognition task aims to recover the text transcription accurately, which is realized by the Deep Speech 2 model~\cite{pmlr-v48-amodei16}.
\subsubsection{\textbf{Benchmark 2}}
The second benchmark is a conventional system that transmits text, named text transceiver. Particularly, the input speech signals are converted into the text sequence before feeding into the conventional system and the text sequence is recovered at the receiver. The Huffman coding~\cite{4051119} is employed for text source coding in the system, the settings of channel coding and modulation are same as that in~\emph{benchmark 1}. In addition, The Deep Speech 2 model is utilized to implement speech recognition at the transmitter. Furthermore, the recovered text sequence is passed through the Tacotron 2 model~\cite{8461368} to reconstruct the speech signals in the speech synthesis task.
\subsubsection{\textbf{Benchmark 3}}
The third benchmark is a conventional system that transmits the extracted text-related semantic features produced by the semantic encoder of DeepSC-ST, named feature transceiver. Those features are the floating-point vectors and encoded to the bit sequence for transmission after passing through the IEEE 754 floating-point arithmetic~\cite{8766229} module, polar code module, and 64-QAM modulator. The text transcription is estimated based on the recovered features at the receiver by leveraging the greedy decoder and the speech signals are reconstructed by feeding the recognized text into the Tacotron 2 model.
\renewcommand\arraystretch{1.15} 
\begin{table}[tbp]
\footnotesize
\caption{Parameter settings for speech recognition task.}
\label{DeepSC-ST NN parameters}
\centering
\begin{tabular}{|c|c|c|c|}
\hline
               & \textbf{Layer Name}  & \textbf{Filters/Units}  & \textbf{Activation}    \\
\hline
            \textbf{Semantic}        &   2$\times$CNN modules   & 2$\times$32   & ReLU \\
\cline{2-4}
            \textbf{Encoder}         &  7$\times$BRNN modules  & 7$\times$800  & Tanh \\
\hline
            \textbf{Channel}        &        Dense layer        &    40         & ReLU \\
\cline{2-4}
            \textbf{Encoder}        &        Dense layer        &    40         & None \\
\hline            
   \multirow{3}{3.5em}{\textbf{\centering Channel\\Decoder}}  &  Dense layer  &  40  &  ReLU \\
\cline{2-4}
                                                              &  Dense layer  &  40  &  ReLU \\
\cline{2-4}
                                                              &  Dense layer  &  29  &  None \\
\hline
\end{tabular}
\end{table}
\subsubsection{\textbf{Benchmark 4}}
The fourth benchmark is a hybrid modeling system that incorporates the speech recognition model and the DeepSC~\cite{9398576} model, named SR+DeepSC. Particularly, the input speech is first converted to the text by the Deep Speech 2 model at the transmitter. Then the text is fed into the DeepSC transmitter before transmission and restored by the DeepSC receiver. Moreover, the recovered text is utilized to synthesize the speech sequence according to the Tacotron 2 model.

Note that the conventional communication paradigm is adopted in the speech transceiver, the text transceiver, and the feature transceiver, while the DL-enabled semantic communication paradigm is leveraged in the SR+DeepSC and the proposed DeepSC-ST.

\subsection{Complexity Analysis}
The system complexity of different transmission schemes for the speech recognition task is introduced as follows. The proposed DeepSC-ST includes 85,796,042 trainable parameters. However, the number of trainable parameters in the SR+DeepSC is 92,212,016, which results in a 7.48$\%$ increase of system complexity over the DeepSC-ST. The system complexity of the feature transceiver is at the same level as the DeepSC-ST. Moreover, the adopted Deep Speech 2 model in the speech transceiver and the text transceiver has 85,788,733 trainable parameters, which nearly mitigates no computational burden than the DeepSC-ST, besides, the extra communication resources are required because of the conventional encoding/decoding mechanism. Therefore, in terms of the system complexity, the proposed DeepSC-ST is at the same level as the speech transceiver, the text transceiver, and the feature transceiver, while slightly lower than the SR+DeepSC.

Furthermore, as aforementioned that the semantic communication system transmits much less amount of data than the source information. We compute the average encoded symbols to transmit 16,000 original speech samples in the proposed DeepSC-ST and the benchmarks, which is summarized in Table~\ref{symbols comparison}. From the table, the proposed DeepSC-ST reduces the transmission data by nearly ten times of the speech transceiver and the feature transceiver while the text transceiver and the SR+DeepSC need average 60 encoded symbols because it only comprises few tokens in the long speech sample sequence.

\subsection{Experiments for Speech Recognition Task}
\begin{figure}[tbp]
\includegraphics[width=0.45\textwidth]{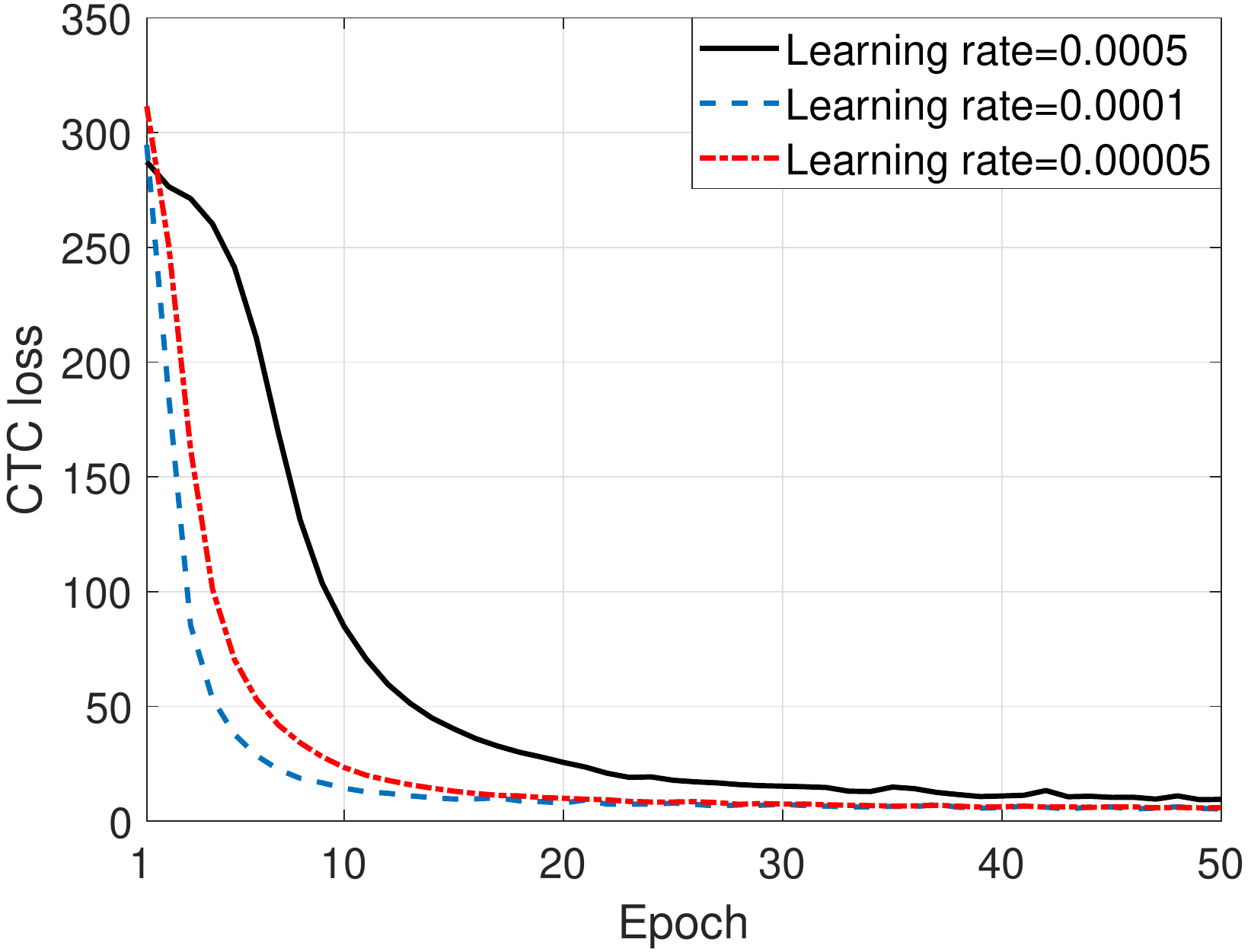}
\centering 
\caption{The training CTC loss versus epoch.}  
\label{Figure: CTC loss VS epoch}  
\end{figure}

The relationship between the CTC loss and the number of epochs is shown in Fig.~\ref{Figure: CTC loss VS epoch}. From the figure, when learning rate is 0.0005, the CTC loss converges slowest and has the highest value amongst the three learning rates, which experiences some fluctuations when epoch$>$30. When learning rate is 0.0001 or 0.00005, the CTC loss decreases to around 10 after 20 epochs and converges after about 40 epochs.

According to (\ref{channel}), the fading channel, $\boldsymbol h$, and the Gaussian noise, $\boldsymbol w$, are specified during the training process and the trained DeepSC-ST is utilized to test the system performance under various channel environments. It is logical that when testing the system performance under a certain channel condition, the DeepSC-ST model trained under the same channel condition performs better than the DeepSC-ST models trained under other channel conditions. However, it is impractical to deploy numerous DeepSC-ST models corresponding to dynamic channel environments because of the scarce computation resources, which inspires us to investigate a DeepSC-ST model that is capable of dealing with channel variations. By comparing the testing performance of different DeepSC-ST models trained under three fading channels, i.e., the AWGN channels, the Rayleigh channels, and the Rician channels, the DeepSC-ST trained under the Rician channels and the Gaussian noise with SNR$=$8 dB is identified as a robust model to cope with diverse channel environments.
\renewcommand\arraystretch{1.15} 
\begin{table*}[tbp]
\footnotesize
\caption{Comparison of encoded symbols in different systems.}
\label{symbols comparison}
\centering
\begin{tabular}{|c|c|c|}
\hline
                        &   Original Speech Samples   &       Encoded Symbols \\
\hline
    DeepSC-ST           &           16,000            &              520          \\
\hline
    Speech Transceiver  &           16,000            &             6,600         \\
\hline
    Feature Transceiver &           16,000            &             6,264         \\
\hline
    Text Transceiver    &           16,000            &              60           \\
\hline
    SR+DeepSC           &           16,000            &              120           \\
\hline
\end{tabular}
\end{table*}
\begin{figure*}[tbp]
\begin{minipage}[t]{0.33\linewidth}
\centering
\includegraphics[width=1\textwidth]{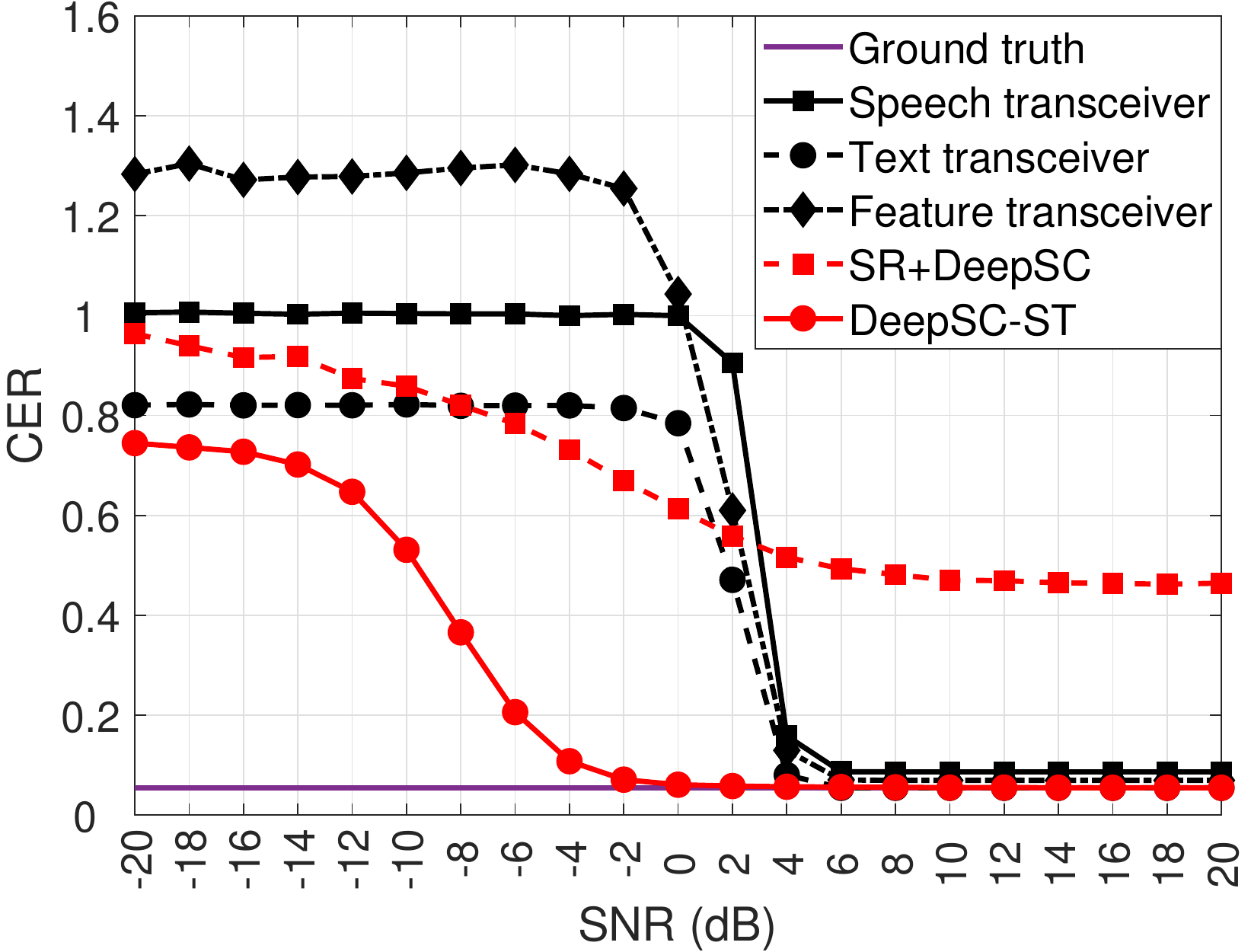}
\subcaption{AWGN channels}
\label{CER AWGN}
\end{minipage}
\begin{minipage}[t]{0.33\linewidth}
\centering
\includegraphics[width=1\textwidth]{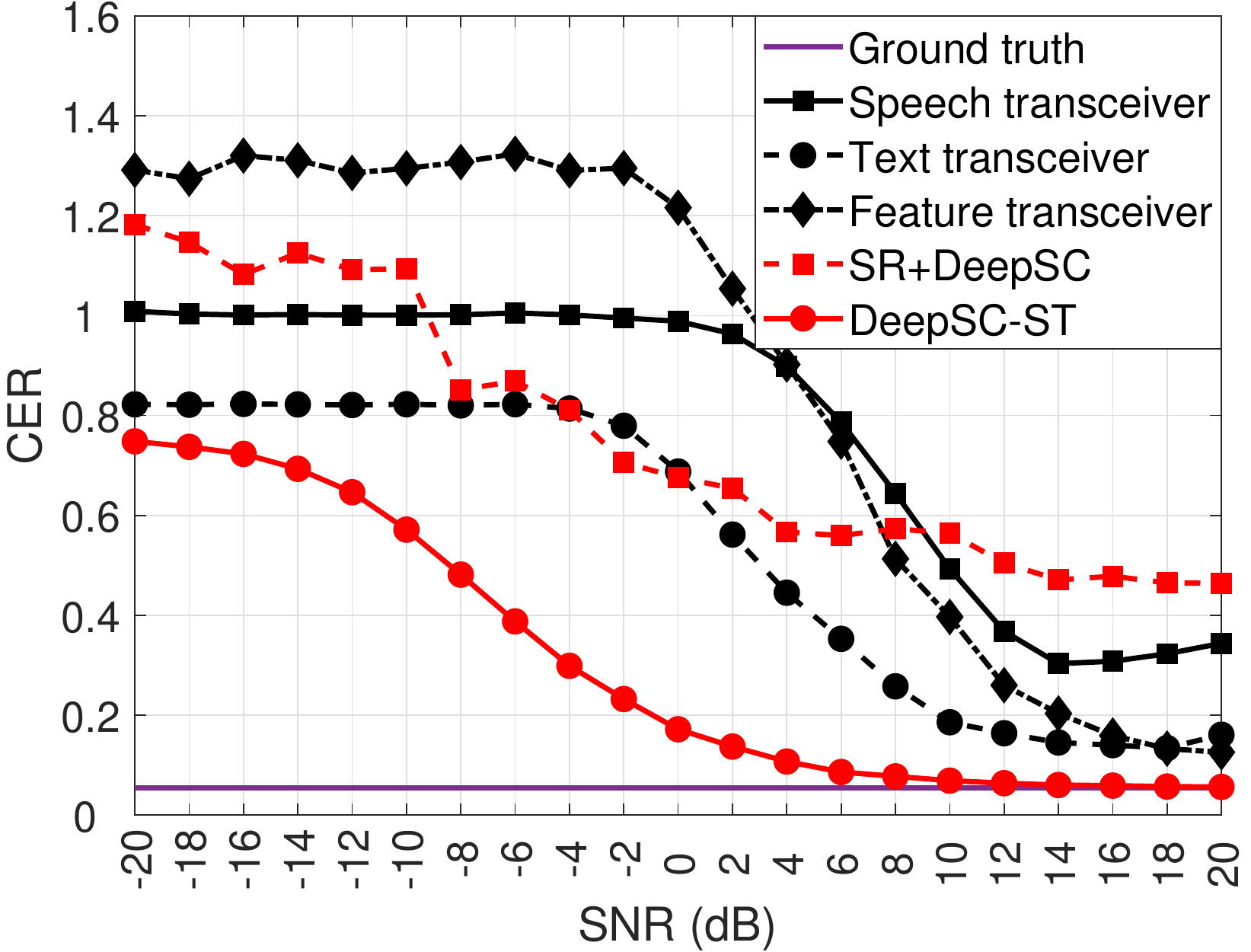}
\subcaption{Rayleigh channels}
\label{CER Rayleigh}
\end{minipage} 
\begin{minipage}[t]{0.33\linewidth}
\centering
\includegraphics[width=1\textwidth]{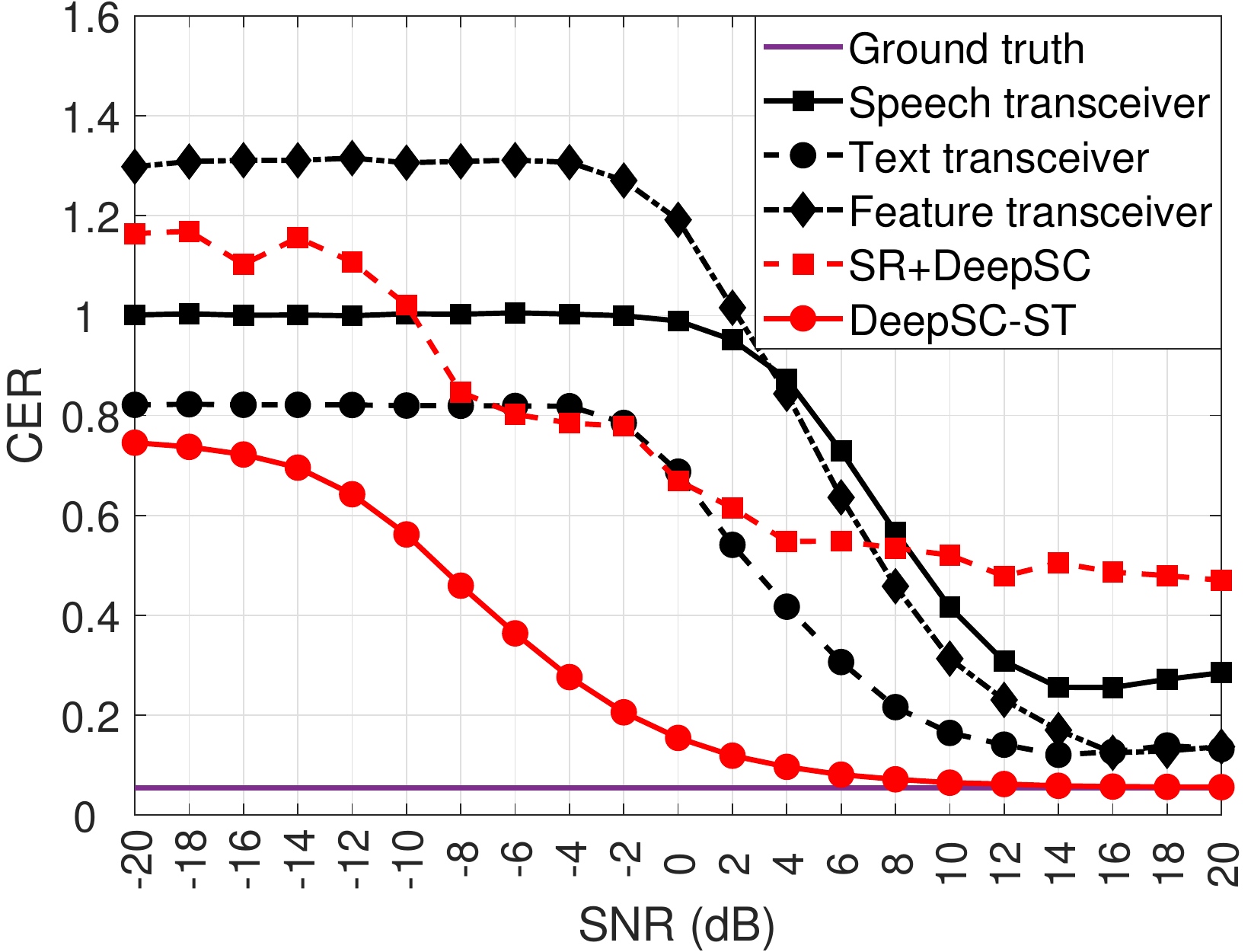}
\subcaption{Rician channels}
\label{CER Rician}
\end{minipage} 
\caption{CER score versus SNR for the speech transceiver, the text transceiver, the feature transceiver, the SR+DeepSC, and the proposed DeepSC-ST.}
\label{CER result}
\end{figure*}
\begin{figure*}[tbp]
\begin{minipage}[t]{0.33\linewidth}
\centering
\includegraphics[width=1\textwidth]{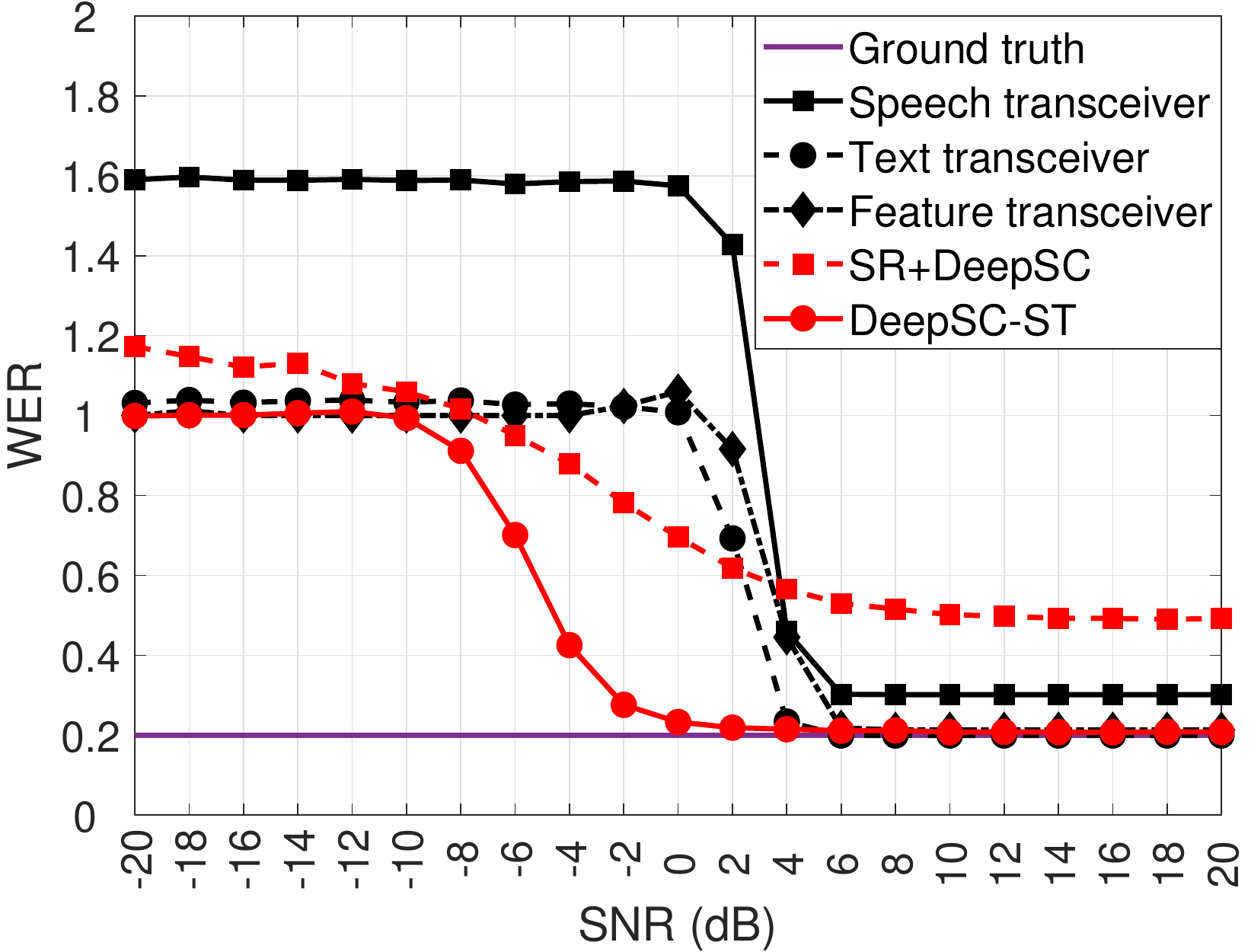}
\subcaption{AWGN channels}
\label{WER AWGN}
\end{minipage}
\begin{minipage}[t]{0.33\linewidth}
\centering
\includegraphics[width=1\textwidth]{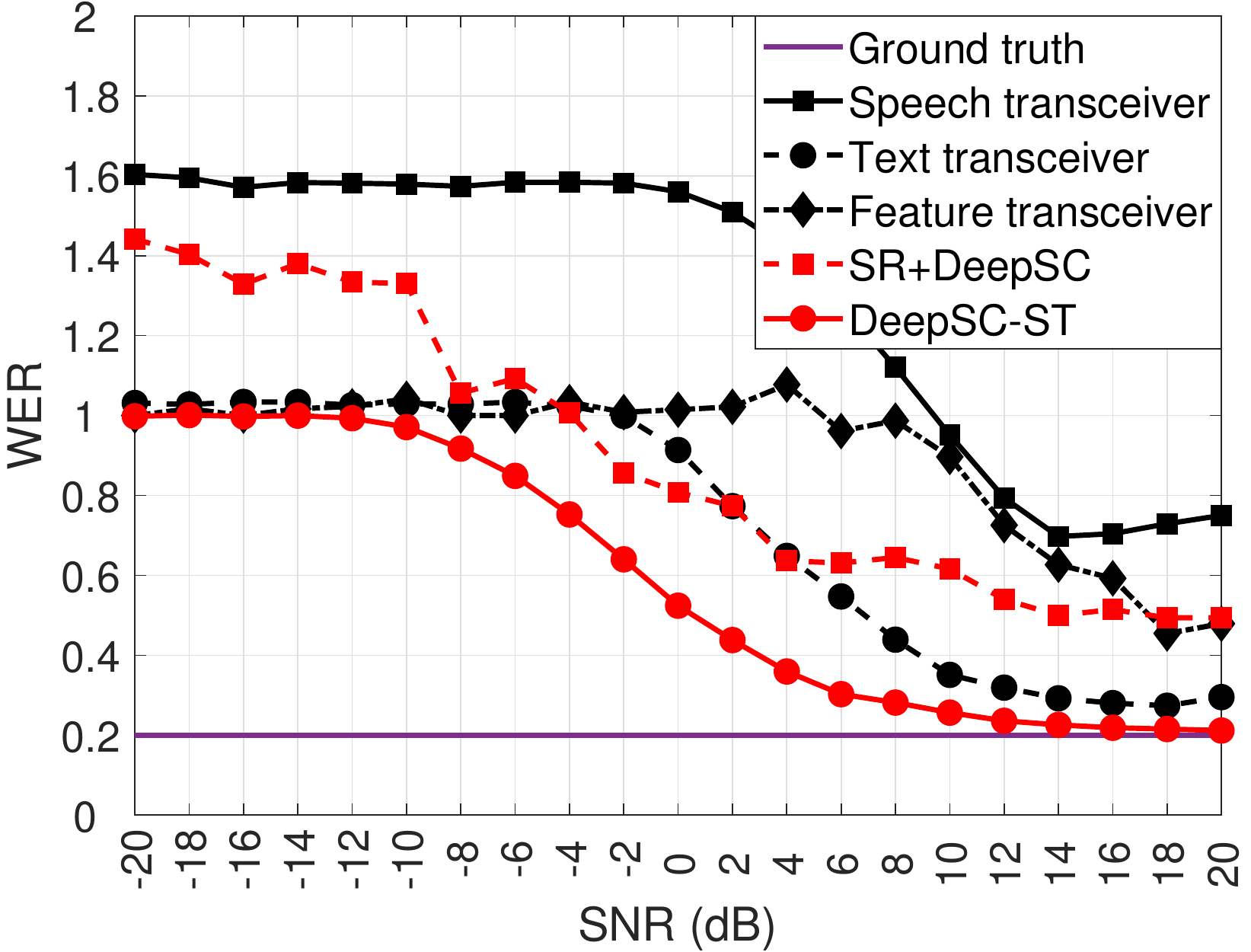}
\subcaption{Rayleigh channels}
\label{WER Rayleigh}
\end{minipage} 
\begin{minipage}[t]{0.33\linewidth}
\centering
\includegraphics[width=1\textwidth]{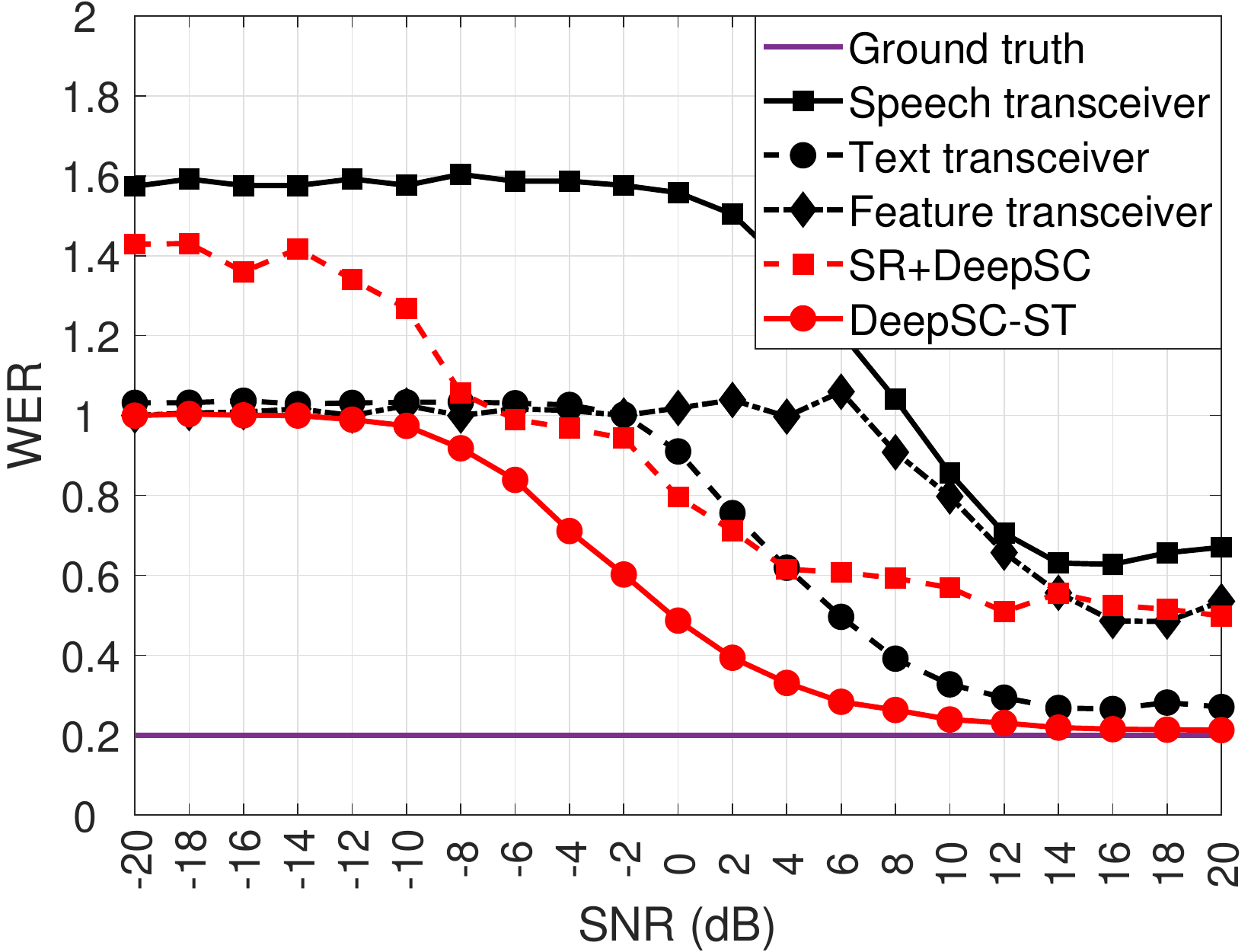}
\subcaption{Rician channels}
\label{WER Rician}
\end{minipage} 
\caption{WER score versus SNR for the speech transceiver, the text transceiver, the feature transceiver, the SR+DeepSC, and the proposed DeepSC-ST.}
\label{WER result}
\end{figure*}
\begin{figure*}[tbp]
\begin{minipage}[t]{0.33\linewidth}
\centering
\includegraphics[width=1\textwidth]{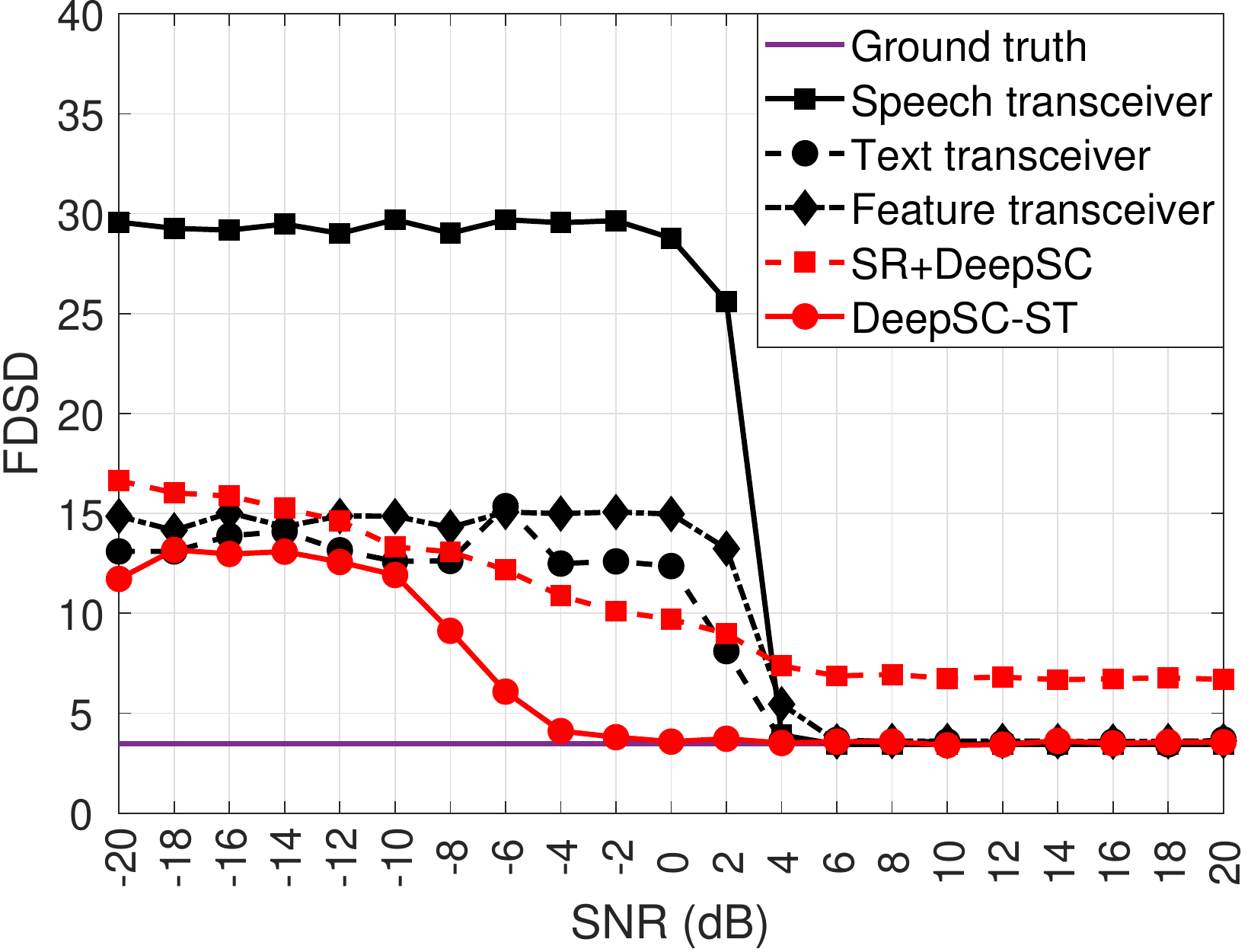}
\subcaption{AWGN channels}
\label{FDSD AWGN}
\end{minipage}
\begin{minipage}[t]{0.33\linewidth}
\centering
\includegraphics[width=1\textwidth]{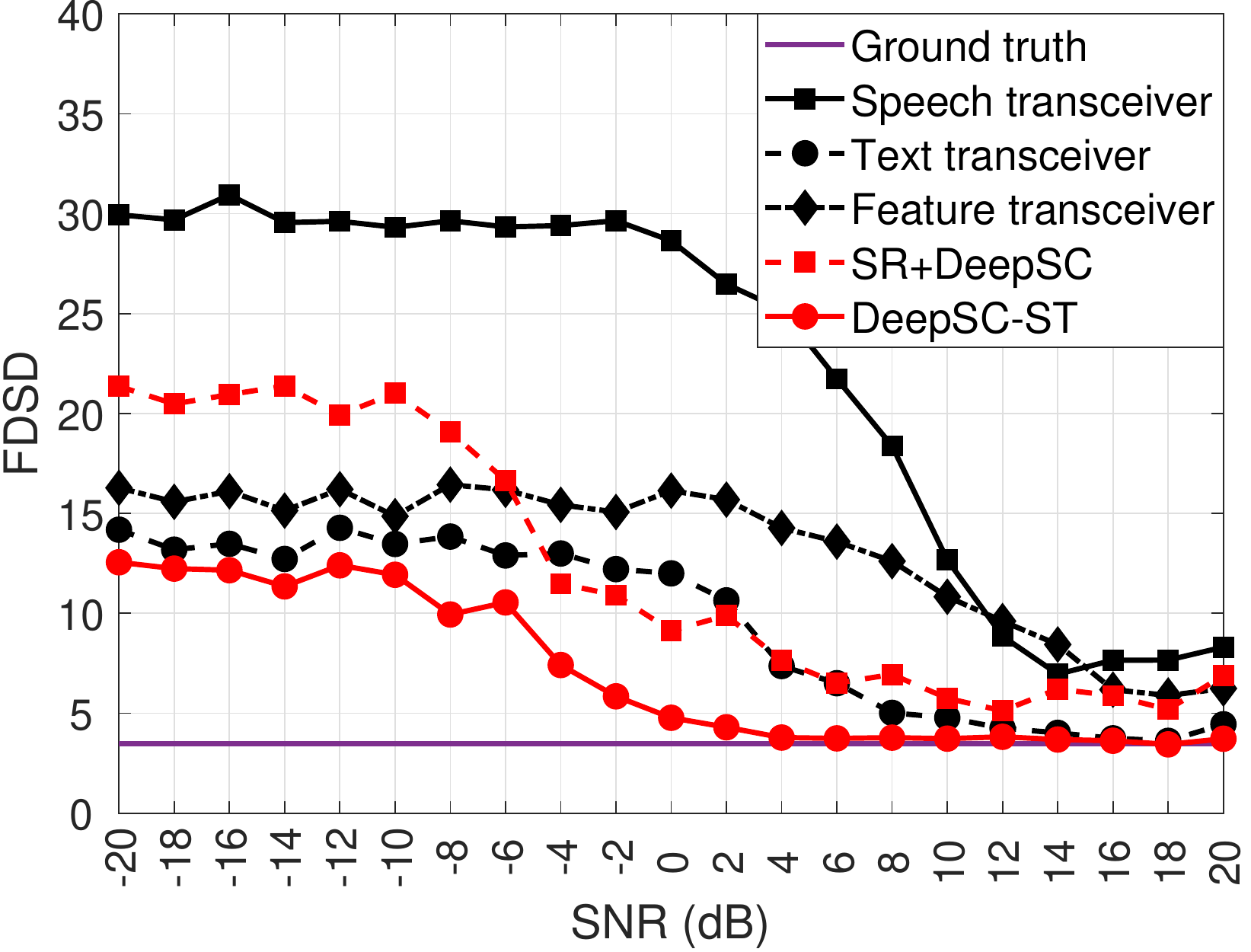}
\subcaption{Rayleigh channels}
\label{FDSD Rayleigh}
\end{minipage} 
\begin{minipage}[t]{0.33\linewidth}
\centering
\includegraphics[width=1\textwidth]{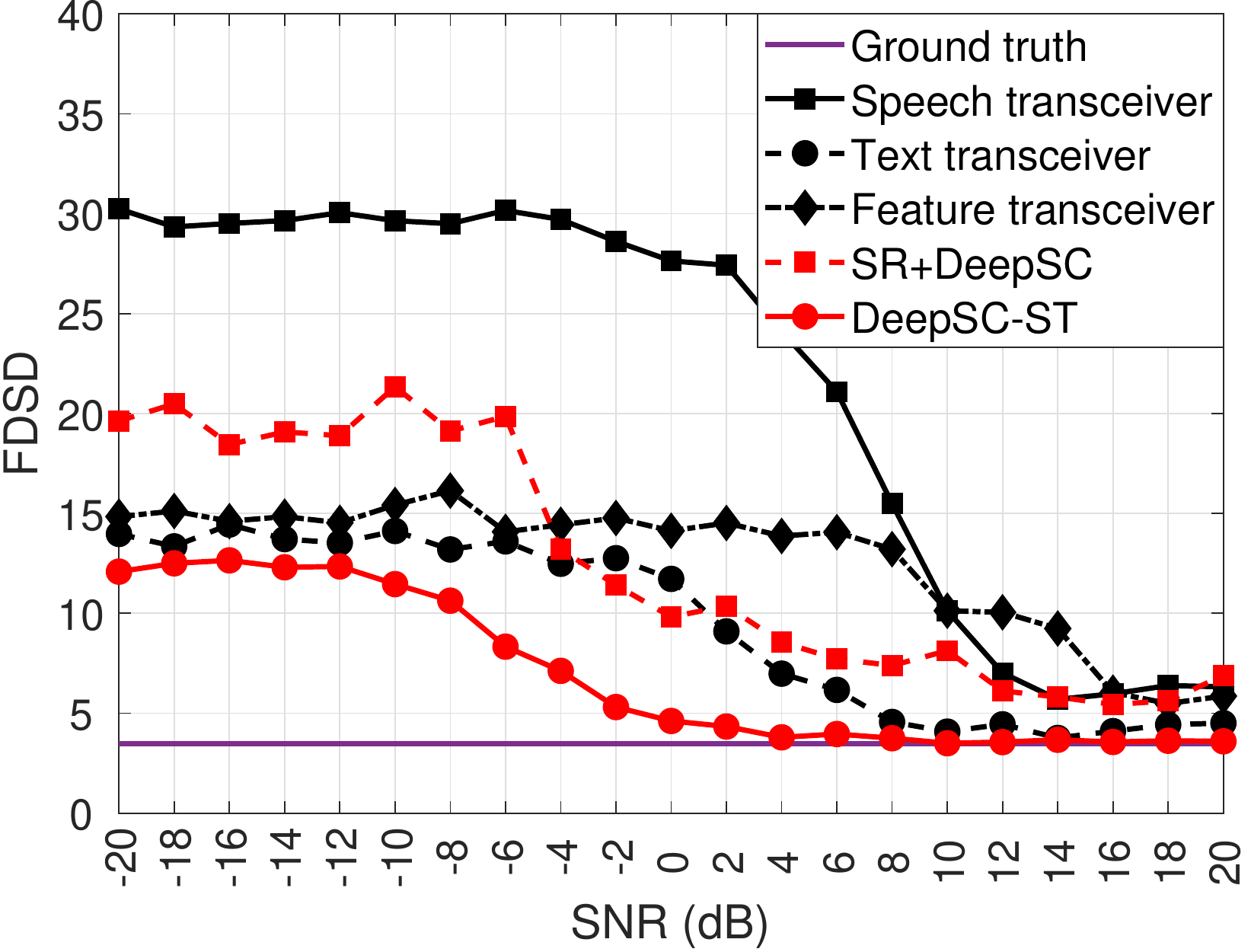}
\subcaption{Rician channels}
\label{FDSD Rician}
\end{minipage} 
\caption{FDSD score versus SNR for the speech transceiver, the text transceiver, the feature transceiver, the SR+DeepSC, and the proposed DeepSC-ST.}
\label{FDSD result}
\end{figure*}
\begin{figure*}[tbp]
\begin{minipage}[t]{0.33\linewidth}
\centering
\includegraphics[width=1\textwidth]{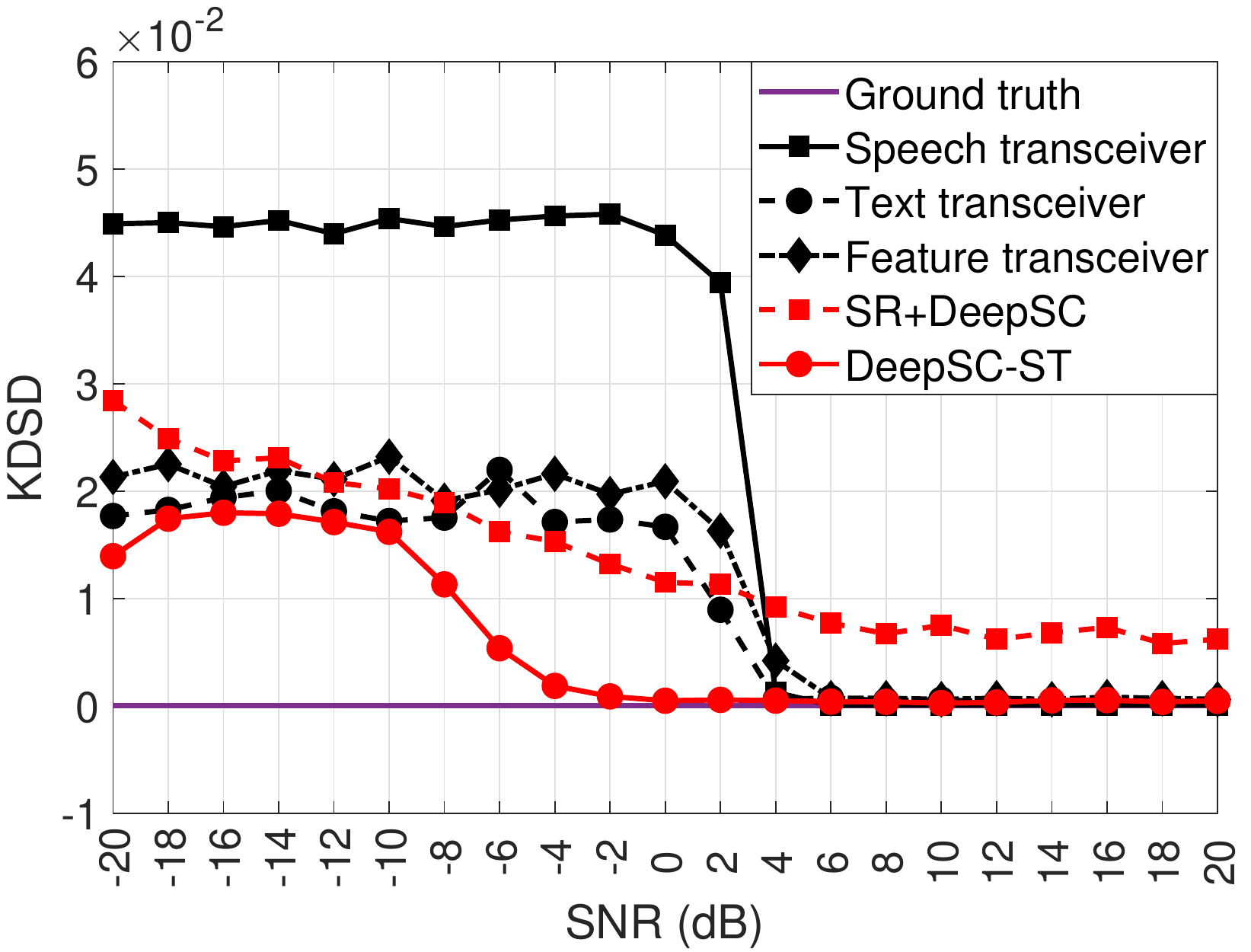}
\subcaption{AWGN channels}
\label{KDSD AWGN}
\end{minipage}
\begin{minipage}[t]{0.33\linewidth}
\centering
\includegraphics[width=1\textwidth]{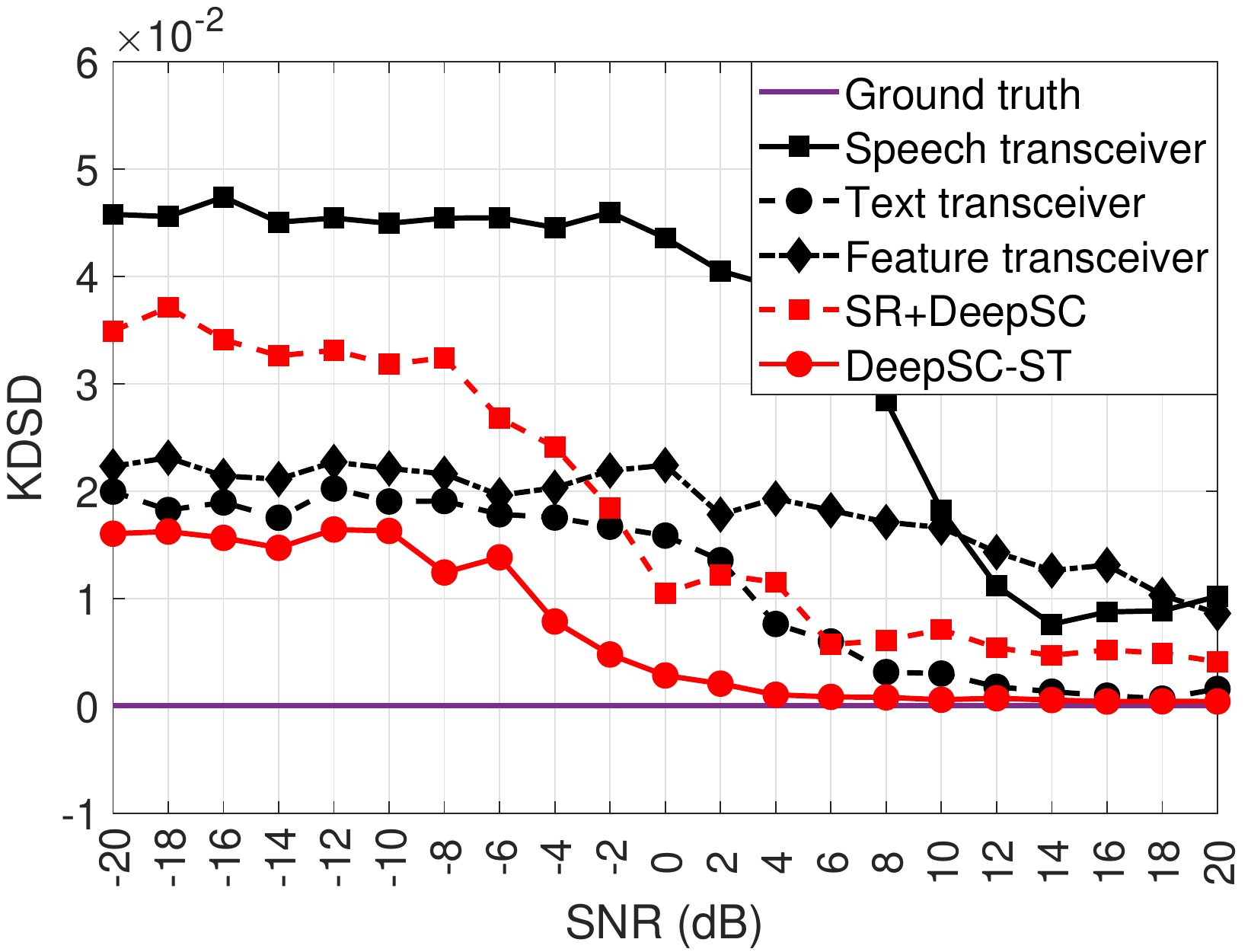}
\subcaption{Rayleigh channels}
\label{KDSD Rayleigh}
\end{minipage} 
\begin{minipage}[t]{0.33\linewidth}
\centering
\includegraphics[width=1\textwidth]{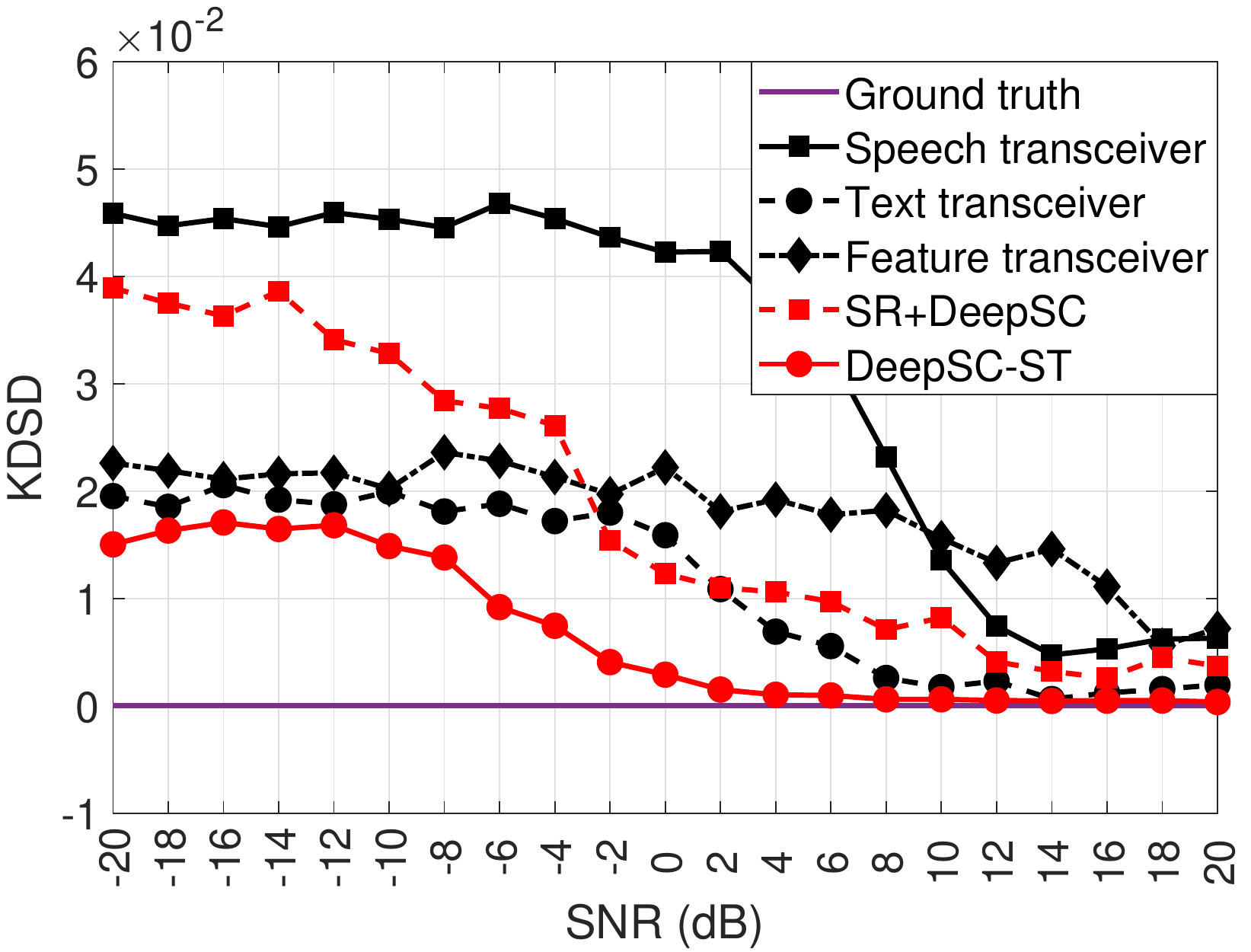}
\subcaption{Rician channels}
\label{KDSD Rician}
\end{minipage} 
\caption{KDSD score versus SNR for the speech transceiver, the text transceiver, the feature transceiver, the SR+DeepSC, and the proposed DeepSC-ST.}
\label{KDSD result}
\end{figure*}

Fig.~\ref{CER result} compares the CER of the DeepSC-ST and four benchmarks for the AWGN channels, the Rayleigh channels, and the Rician channels, where the ground truth is the result tested by feeding the speech sample sequence into the Deep Speech 2 model directly without considering communication problems. From the figure, the DeepSC-ST has lower CER scores than the benchmarks under all tested channel environments. Particularly, the recognized text transcription in the proposed DeepSC-ST is readable when SNR is higher than -2 dB under the AWGN channels while the required SNR is 4 dB in the speech transceiver, the feature transceiver, and the text transceiver. SR+DeepSC obtains no readable text under the adopted fading channels and SNRs. In addition, the DeepSC-ST performs steadily when coping with dynamic channels and SNR$>$0 dB. However, the performance of the benchmarks is quite poor under different channel conditions. Moreover, the DeepSC-ST significantly outperforms the benchmarks when the SNR ranges from -12 dB to 4 dB for the AWGN channels, and -12 dB to 8 dB for the Rayleigh channels and the Rician channels. The fluctuation in the speech transceiver and the text transceiver for SNR$>$14 dB is because the Deep Speech 2 model is trained under the clear speech signals, which results in the slight uncertainty to process the speech signals at the similar noise level.

Fig.~\ref{WER result} compares the WER of different approaches. From the figure, the proposed DeepSC-ST provides lower WER and outperforms the speech transceiver under various channel conditions, as well as the text transceiver and the feature transceiver when SNR$<$8 dB. Moreover, similar to the results of CER, the DeepSC-ST has low WER on average when coping with channel variations while the conventional systems and SR+DeepSC provide poor WER scores when SNR is low. According to the simulation results, the DeepSC-ST is able to achieve better performance to recover the text transcription at the receiver from the input speech signals at the transmitter when coping with the complicated communication scenarios than the conventional communication systems and semantic communication systems, especially in the low SNR regime.

\subsection{Experiments for Speech Synthesis Task}
In this experiment, we test FDSD and KDSD scores of the received speech in the speech transceiver, the synthesized speech in the feature transceiver, the text transceiver, the SR+DeepSC, and the proposed DeepSC-ST. In practice, the audio is generally treated as unacceptable when it is unable to understand or full of background noise. Therefore, FDSD and KDSD thresholds are necessary to determine the validity of the audio. Inspired by this, a survey is developed to collect the opinions of 100 evaluators by selecting the satisfaction degree of the speech signals corresponding to different FDSD and KDSD scores. Particularly, the Deep Speech 2 model is utilized to extract the aforementioned features, $\boldsymbol D$ and $\widehat{\boldsymbol D}$, to compute FDSD and KDSD scores. The survey result is summarized in Table~\ref{FDSD/KDSD survey result}\footnote{Note that KDSD and FDSD scores are computed under the adopted Deep Speech 2 model in our experiment, these scores may different when the model parameters are varied or other models are employed.}. From the table, the speech signals can be considered as acceptable when their FDSD or KDSD scores are less than 10 or 0.012, respectively.

The FDSD and KDSD results of the DeepSC-ST and four benchmarks are shown in Fig.~\ref{FDSD result} and Fig.~\ref{KDSD result}, respectively, where the ground truth is the KDSD and FDSD scores computed by passing the plain text sequence through the Tacotron 2 model directly. From the figure, the DeepSC-ST obtains lower FDSD and KDSD scores than the SR+DeepSC under all tested channel conditions, besides, it achieves better speech recovery, i.e., lower FDSD and KDSD scores, than the speech transceiver, the feature transceiver, and the text transceiver from -8 dB to 2 dB under the AWGN channels, as well nearly -10 dB to 10 dB under the Rayleigh channels and the Rician channels. Particularly, when SNR is lower than around 4 dB under the AWGN channels, the received speech signals in the speech transceiver are invalid while the synthesized waveform in the proposed DeepSC-ST is acceptable for SNR$>$-10 dB, which proves the adaptability of DeepSC-ST in the low SNR regime. In addition, the proposed DeepSC-ST outperforms the adopted benchmarks under the Rayleigh channels and the Rician channels amongst all the tested SNRs. Furthermore, the DeepSC-ST is robust to cope with the diverse channel environments because the SNR thresholds yielding the valid speech waveform are nearly the same in different fading channels in different fading channels.
\subsection{DeepSC-ST Demonstration}
\begin{figure}[tbp]
\includegraphics[width=0.45\textwidth]{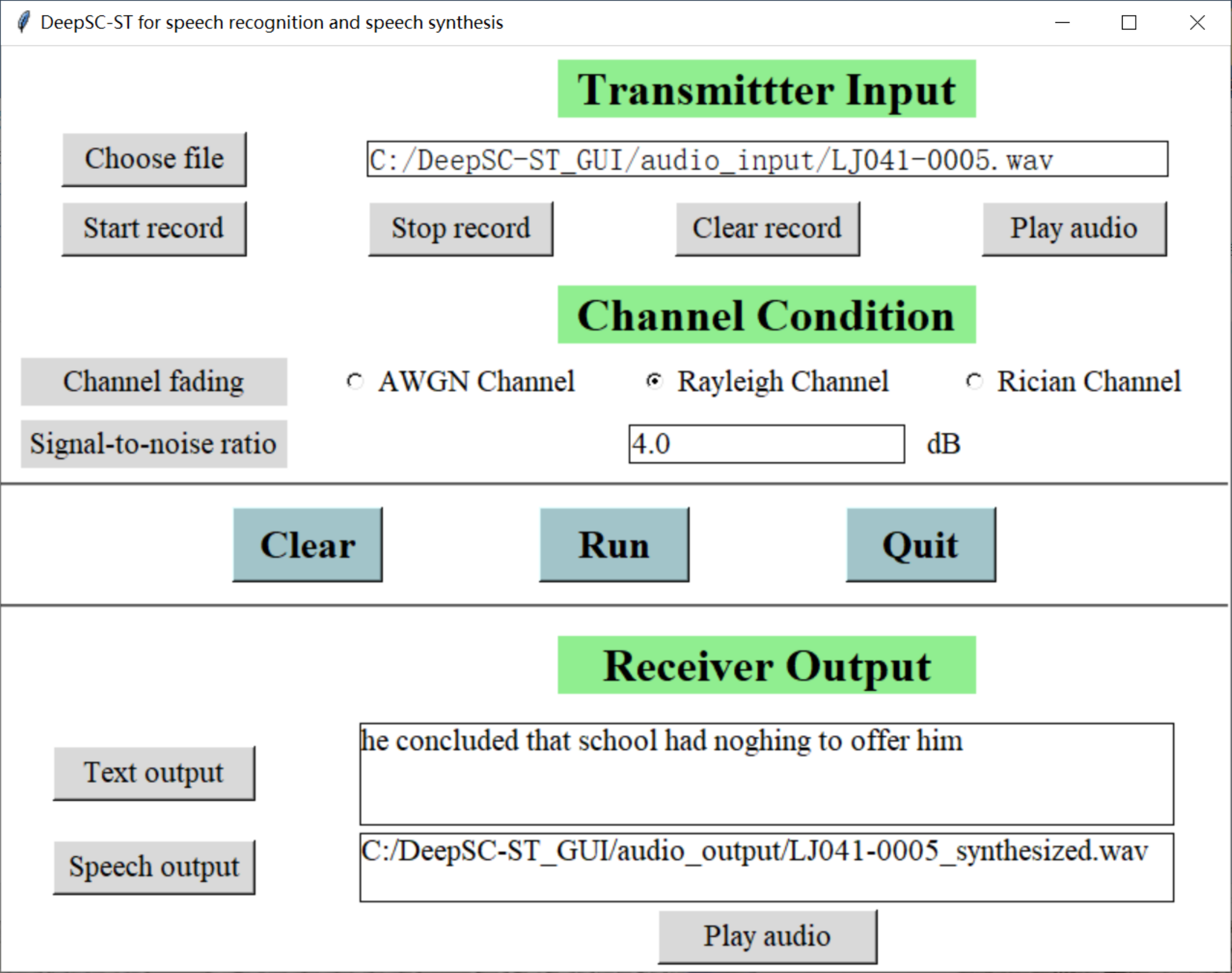}
\centering 
\caption{Software demonstration of the developed DeepSC-ST.}  
\label{Figure: DeepSC-ST demo}
\end{figure}

\begin{figure*}[tbp]
\begin{minipage}[t]{0.5\linewidth}
\flushright
\subcaption{Original}
\includegraphics[width=0.85\textwidth]{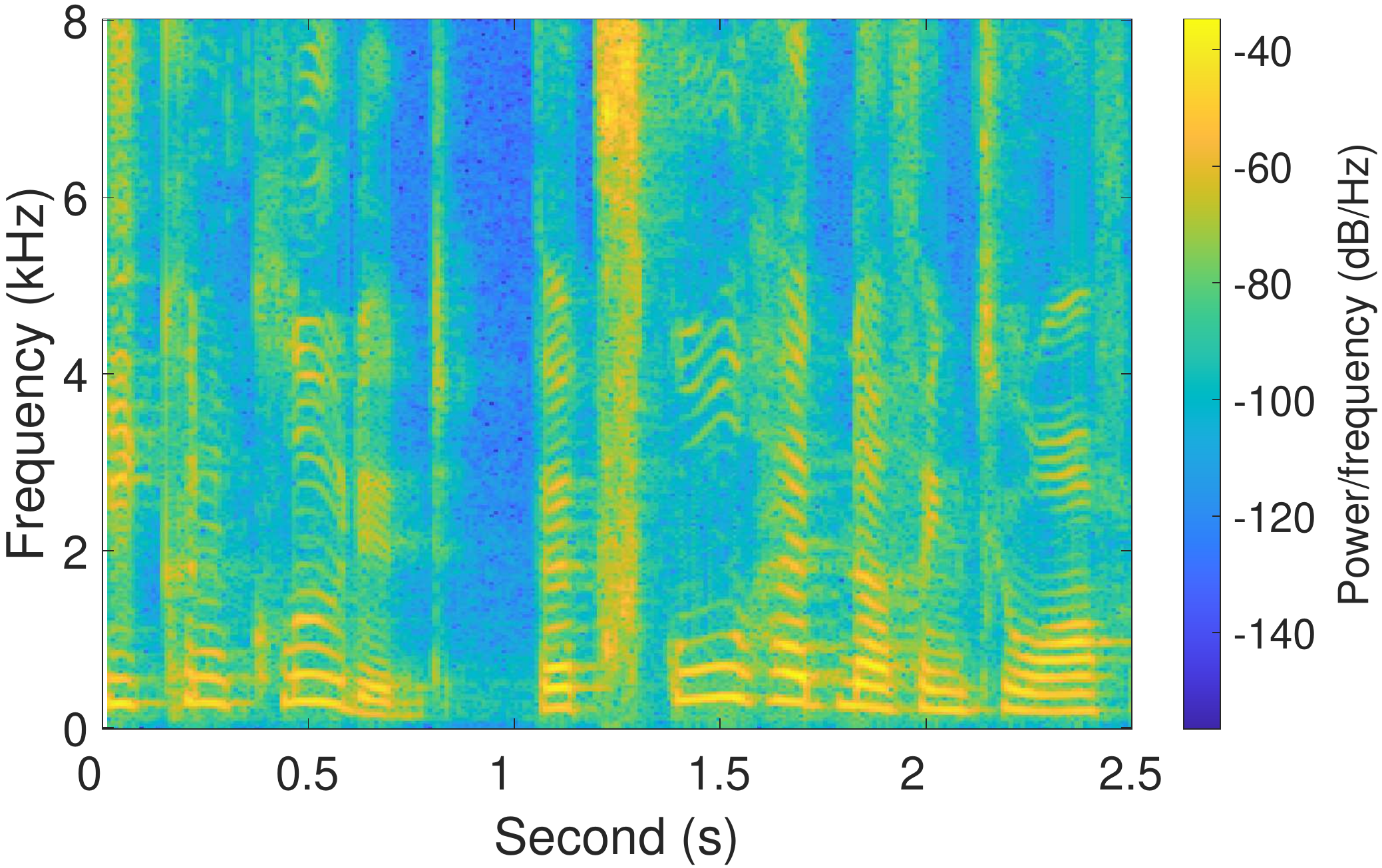}
\label{spectrogram orginal}
\end{minipage}
\begin{minipage}[t]{0.5\linewidth}
\flushleft
\subcaption{DeepSC-ST}
\includegraphics[width=0.85\textwidth]{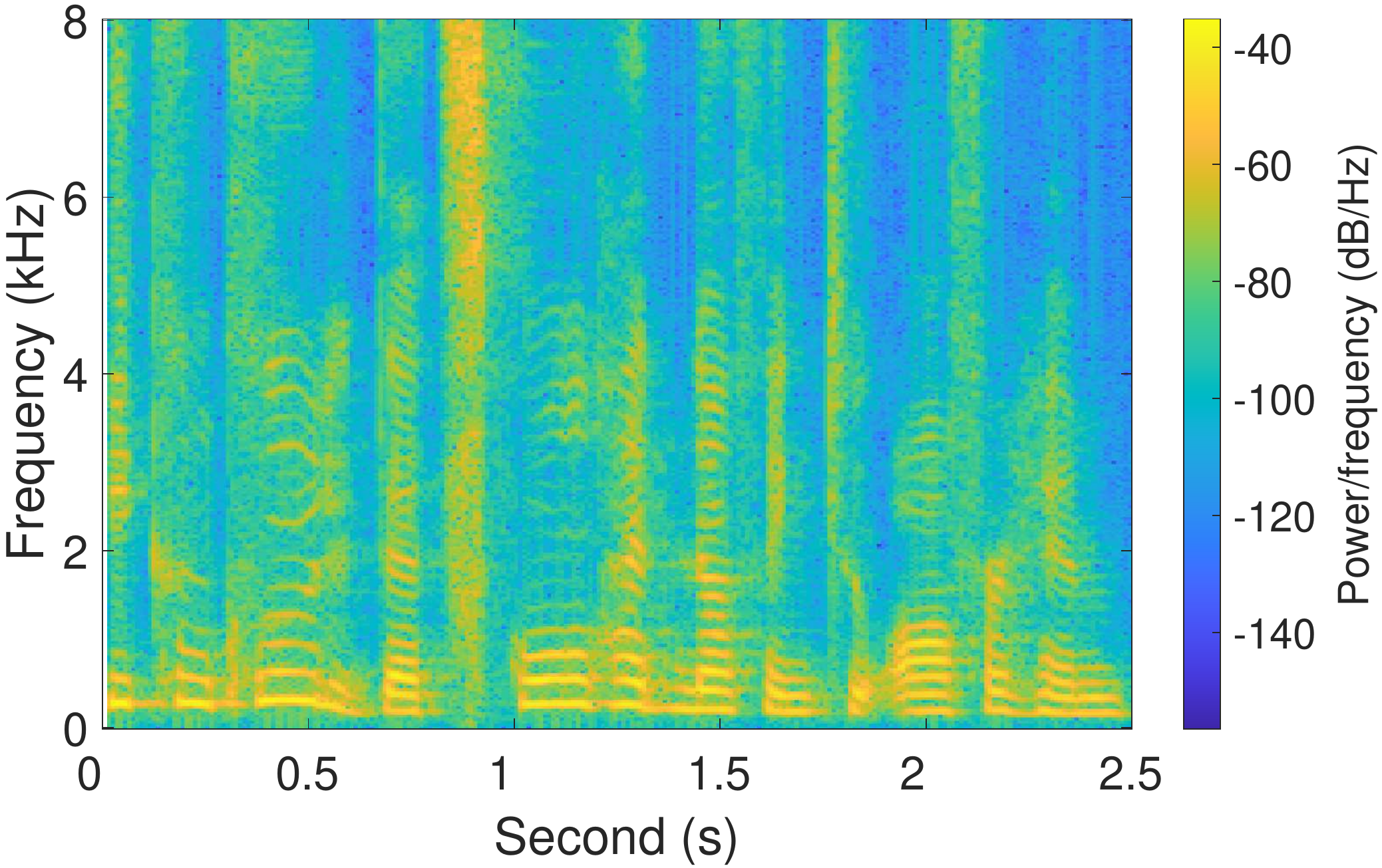}
\label{spectrogram DeepSC-ST}
\end{minipage}
\begin{minipage}[t]{0.5\linewidth}
\flushright
\subcaption{Speech transceiver}
\includegraphics[width=0.85\textwidth]{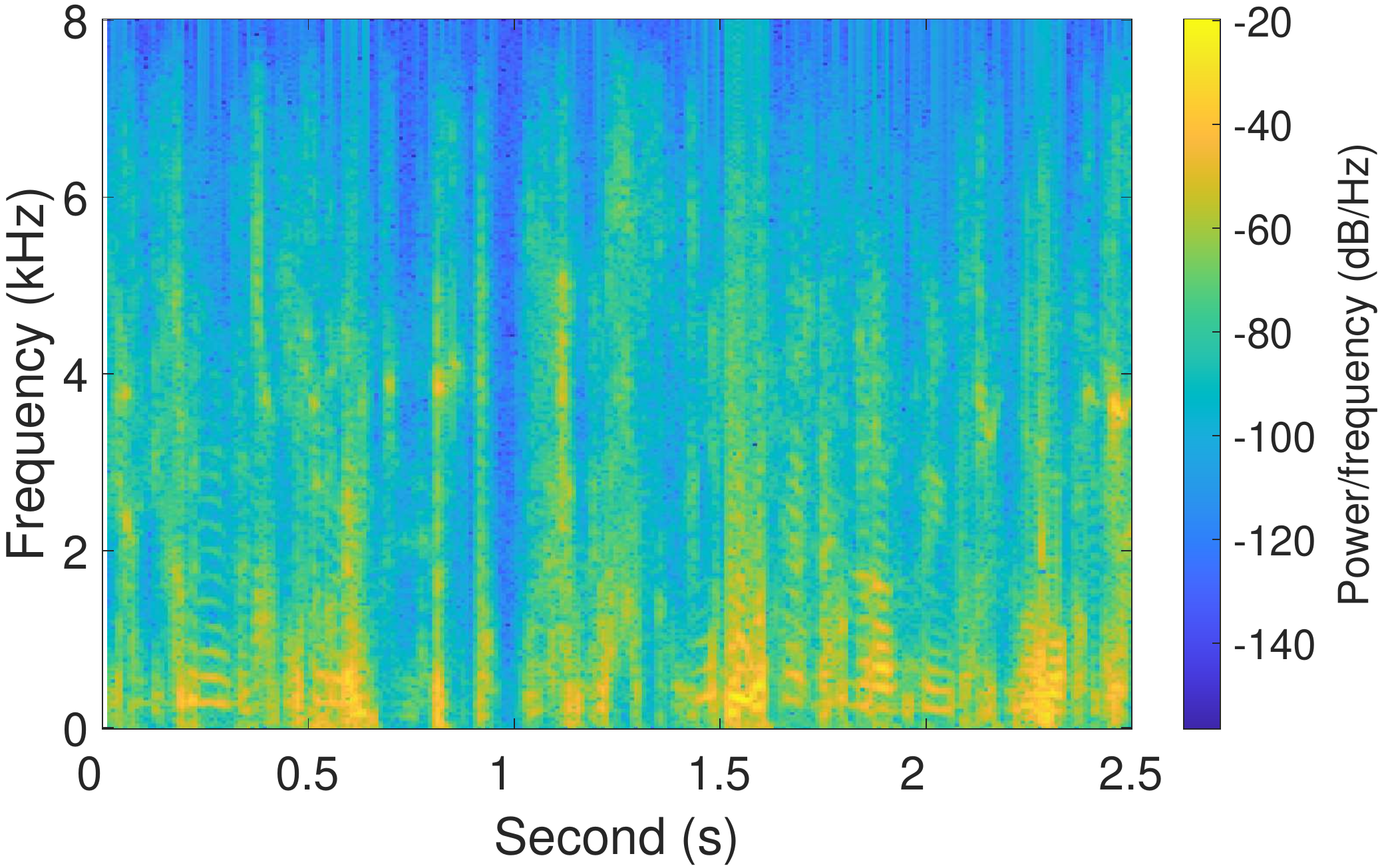}
\label{spectrogram speech transceiver}
\end{minipage}
\begin{minipage}[t]{0.5\linewidth}
\flushleft
\subcaption{Feature transceiver}
\includegraphics[width=0.85\textwidth]{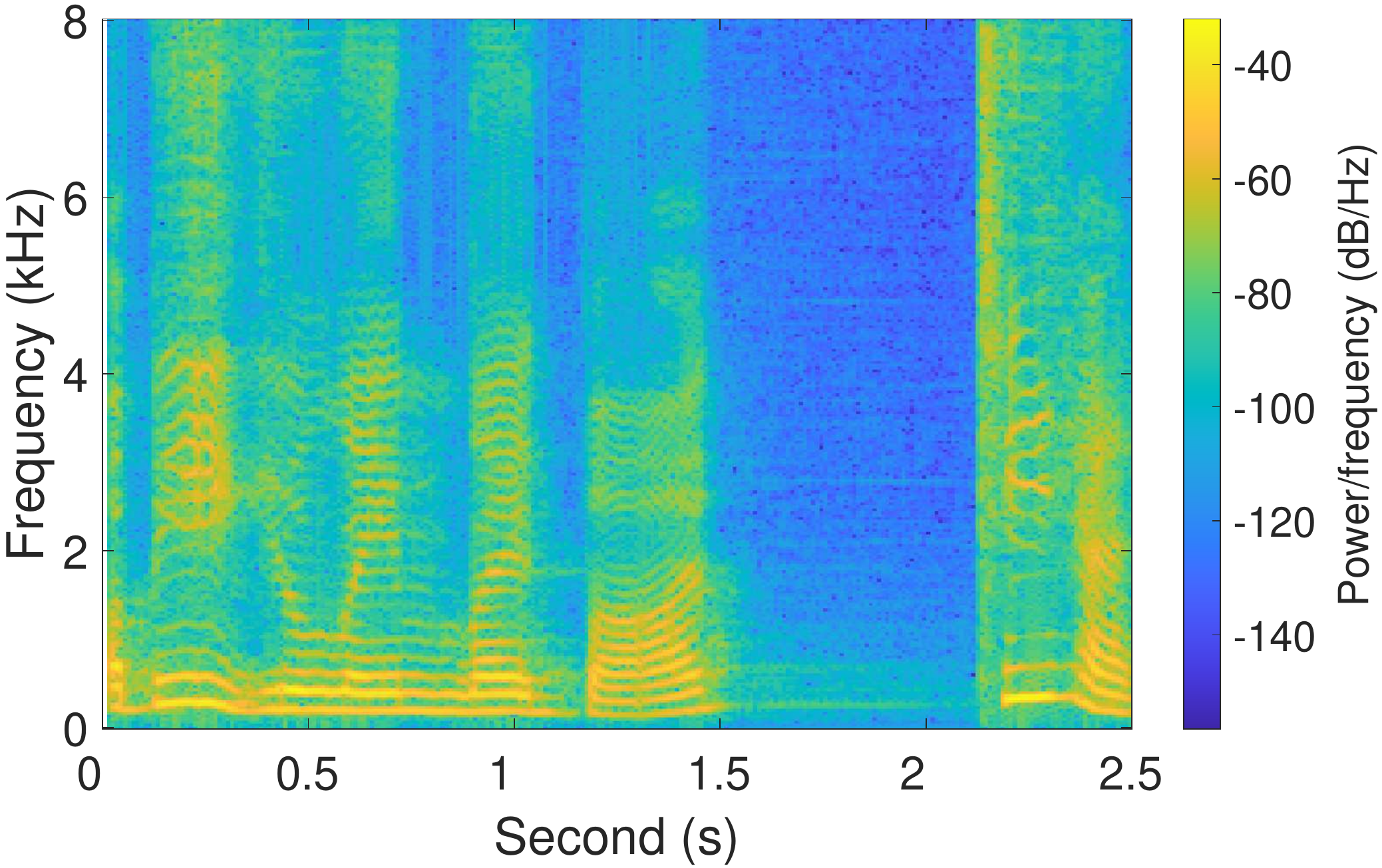}
\label{spectrogram feature transceiver}
\end{minipage}
\begin{minipage}[t]{0.5\linewidth}
\flushright
\subcaption{Text transceiver}
\includegraphics[width=0.85\textwidth]{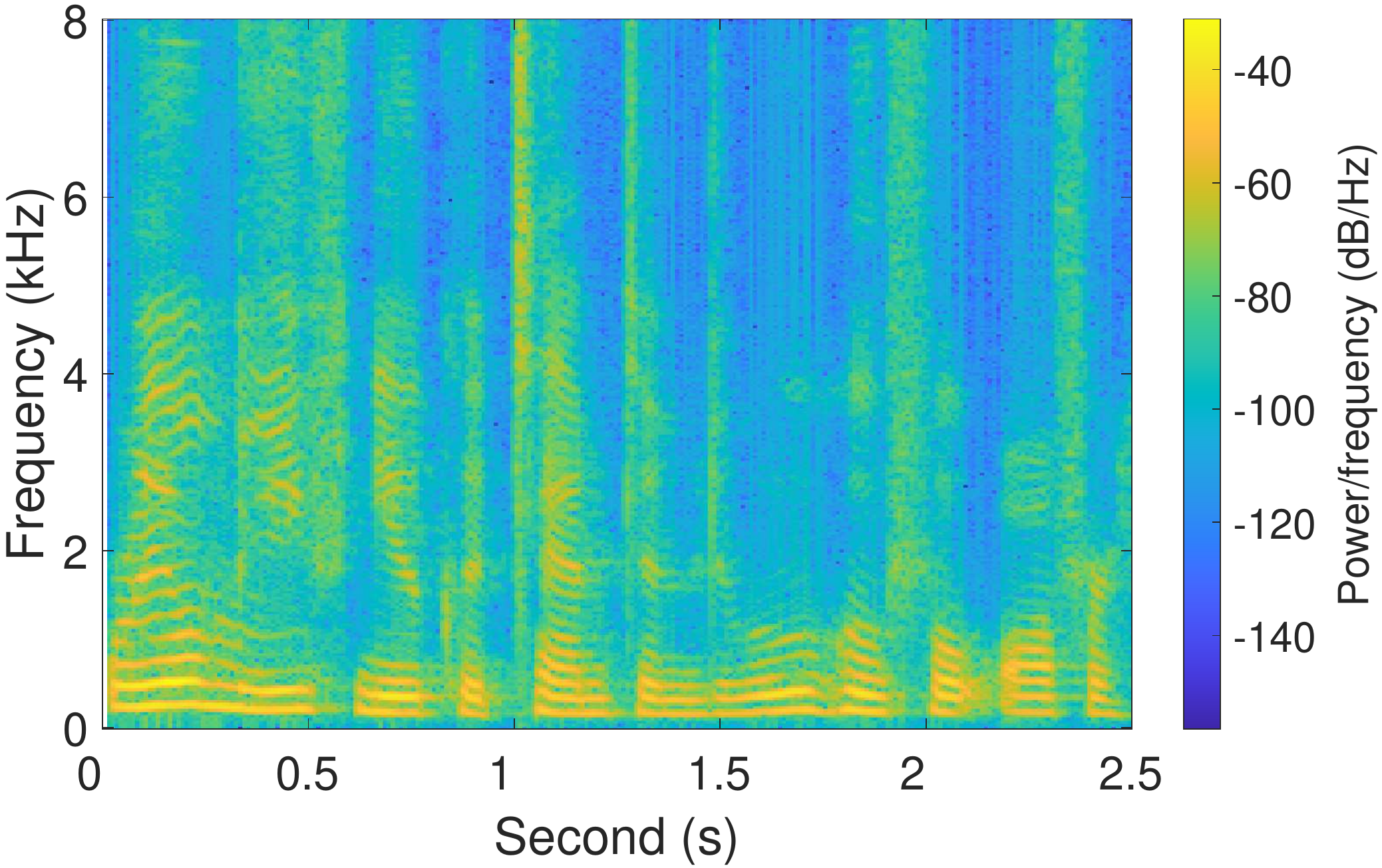}
\label{spectrogram text transceiver}
\end{minipage} 
\begin{minipage}[t]{0.5\linewidth}
\flushleft
\subcaption{SR+DeepSC}
\includegraphics[width=0.85\textwidth]{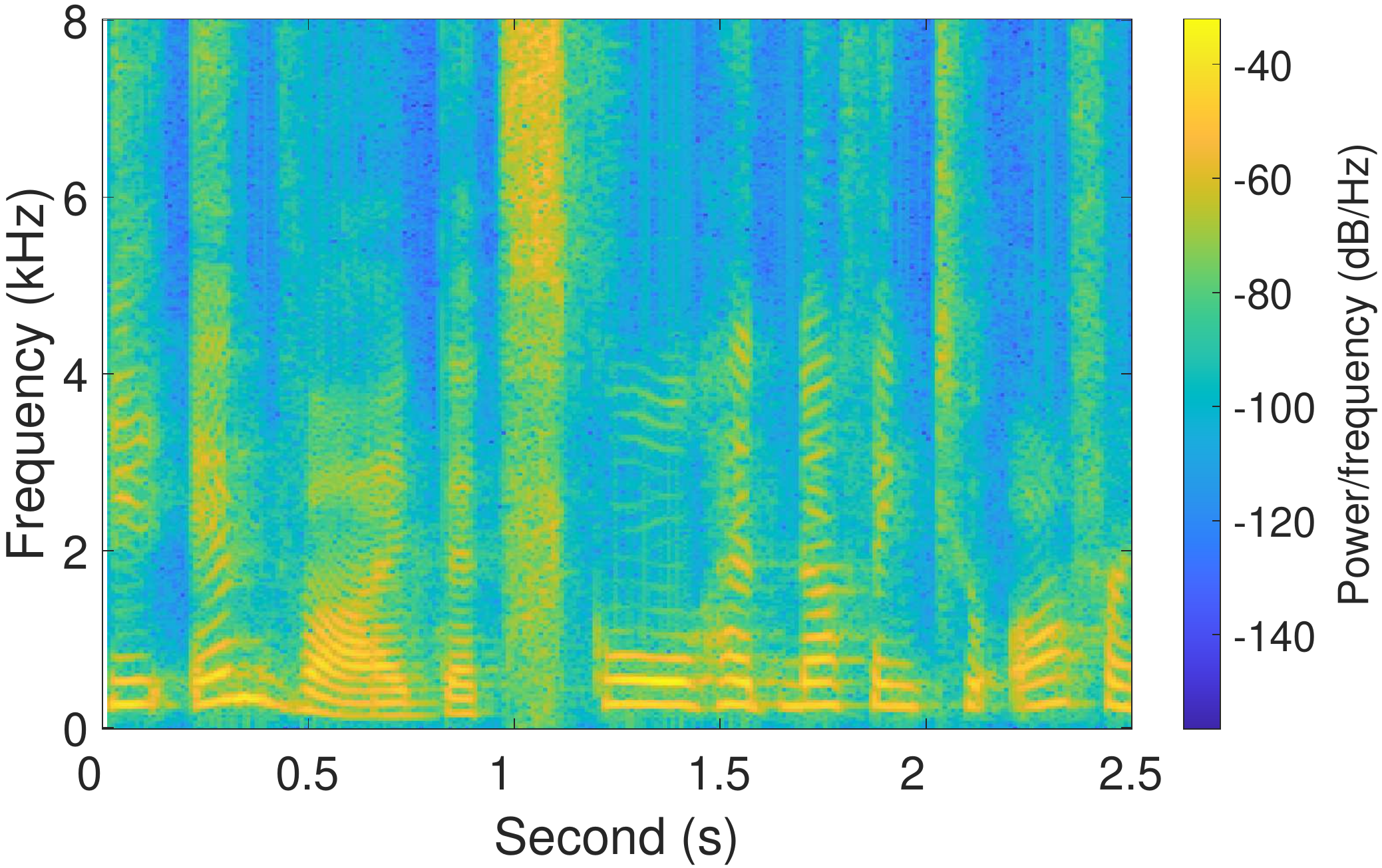}
\label{spectrogram SR+DeepSC}
\end{minipage}
\caption{Spectrograms of speech sample sequences reconstructed by different systems over Rayleigh channel when SNR is 4 dB.}
\label{spectrogram result}
\end{figure*}

According to the proposed DeepSC-ST, we design a software demonstration with user interface, as shown in Fig.~\ref{Figure: DeepSC-ST demo}\footnote{This software demonstration can be found at \url{https://github.com/Zhenzi-Weng/DeepSC-ST_demonstration}.}. In the figure, users could choose a local~\emph{.wav} file or record the speech input. Besides, the software demonstration allows users to specify a fading channel amongst three different channels and input an arbitrary SNR value. Then, the recognized text and the synthesized speech are obtained after running a pre-trained DeepSC-ST model. The representative results of the recognized text and the synthesized speech are shown in Table~\ref{comparison of sentences} and Fig.~\ref{spectrogram result}, respectively.

\section{Conclusions}
\renewcommand\arraystretch{1.15} 
\begin{table}[tbp]
\footnotesize
\caption{FDSD and KDSD survey result.}
\label{FDSD/KDSD survey result}
\centering
\begin{tabular}{|c|c|c|}
\hline
                &    FDSD    &    KDSD         \\
\hline
    Quite Clear &  $\leq2$   & $\leq0.001$     \\
\hline
    Clear       & $2\sim10$  & $0.001\sim0.012$ \\
\hline
    Acceptable  & $10\sim18$ & $0.012\sim0.025$ \\
\hline
    Noisy       & $18\sim30$ & $0.025\sim0.045$ \\
\hline
    Quite Noisy & $\geq30$   & $\geq0.045$ \\
\hline
\end{tabular}
\end{table}
\renewcommand\arraystretch{1.15} 
\begin{table*}[tbp]
\footnotesize
\caption{Recognized sentences in different systems over Rayleigh channel when SNR is 4 dB.}
\label{comparison of sentences}
\centering
\begin{tabular}{|c|c|}
\hline
    Original Sentence   & he concluded that school had nothing to offer him   \\
\hline
    DeepSC-ST           & he concluded that school had noghing to offer him    \\
\hline
    Speech Transceiver  & it cood sili ite bebou a pims t lup ar of ig mote terigytit w \\
\hline
    Feature Transceiver & oba uolaihmlud bod tyhat uschtodhzmax nvilitjpaomgjezeb hatm \\
\hline
    Text Transceiver    & h  ea aahesourhhtntchchen ehaoeitdcofo offer him              \\
\hline
    SR+DeepSC           & he rebuilt that school had nothing to pay him and he \\
\hline
\end{tabular}
\end{table*}

In this paper, we investigated a DL-enabled semantic communication system for speech recognition and speech synthesis tasks, named DeepSC-ST, which recovers the text transcription by utilizing the text-related semantic features and reconstructs the speech sample sequence at the receiver. Particularly, we design a joint semantic-channel coding scheme to learn and extract semantic features and mitigate the channel effects to achieve the speech recognition. Simulation results verified that the DeepSC-ST outperforms the conventional communication systems and the existing semantic communication systems, especially in the low SNR regime. Moreover, we built a software demonstration to allow the real human speech input for the proof-of-concept. Our proposed DeepSC-ST is envisioned to be a promising candidate for semantic communication systems for speech recognition and speech synthesis tasks.

\bibliographystyle{IEEEtran}
\bibliography{reference.bib}

\end{document}